\documentclass[12pt]{article}
\usepackage[left=1in,top=1in,right=1in,bottom=1in]{geometry}

\usepackage{geometry,setspace}

\usepackage{natbib}

\usepackage[hidelinks]{hyperref}
\usepackage{optidef}

\usepackage{amsmath}% Advanced Math Typesetting
\usepackage{amsfonts}% Unicode support
\usepackage{amssymb}
\usepackage{makeidx}
\usepackage{graphicx}% Add pictures to document
\usepackage{subcaption}

\usepackage{makeidx, siunitx, ragged2e}
\usepackage{booktabs,color}
\usepackage{amsfonts,bm}
\usepackage{amsmath}
\usepackage{amssymb}
\usepackage{graphicx}
\usepackage{rotating}
\usepackage{multirow}%
\usepackage{multicol}
\usepackage{rotating}
\usepackage{tikz}
\usepackage{pdfpages}
\usepackage{ulem}
\usepackage{booktabs}
\usepackage{lscape}
\usepackage{caption}
\usepackage{lipsum}
\usepackage{longtable}
\usepackage{booktabs}
\newcommand{\rc}{\color{red}}

\newcommand{\nc}{\normalcolor}

\begin{document}
%%%%%%%%%%%%%%%%%%%%%%%%%%%%%%%%%%%%%%%%%%%%%%%%%%%%%%%%%%%%%%%%%%%%%%%%%%%%%%

\title{Bitcoin Volatility and Intrinsic Time Using
Double Subordinated L\'{e}vy Processes}
%: A global dollar Environment index}

\author{Abootaleb Shirvani\thanks{Department of Mathematical Science, Kean University,  ashirvan@kean.edu.}
	\and 
 Stefan Mittnik\thanks{Department of Statistics,  Ludwig Maximilians University, mittnik@gmx.de.}
	\and 
 W. Brent Lindquist\thanks{Department of Mathematics
	\& Statistics, Texas Tech University, brent.lindquist@ttu.edu.}
	\and 
 	Svetlozar Rachev\thanks{Department of Mathematics
	\& Statistics, Texas Tech University, zari.rachev@ttu.edu.}
}
\date{}
\maketitle
%%%%%%%%%%%%%%%%%%%%%%%%%%%%%%%%%%%%%%%%%%%%%%%%%%%%%%%%%%%%%%%%%%%%%%%%%%%%%%%%%%%%%%%%%%%

\begin{abstract}
 We propose a doubly subordinated Levy process, NDIG, to model the time series properties of the cryptocurrency bitcoin.
NDIG captures the skew and fat-tailed properties of bitcoin prices and gives rise to an arbitrage free, option pricing model.
In this framework we derive two bitcoin volatility measures.
The first combines NDIG option pricing with the Cboe VIX model to compute an implied volatility;
the second uses the volatility of the unit time increment of the NDIG model.
Both are compared to a volatility based upon historical standard deviation.
With appropriate linear scaling, the NDIG process perfectly captures observed, in-sample, volatility.\\
(\textit{JEL}: C02, G12, G13)\\

 
\end{abstract}

%%%%%%%%%%%%%%%%%%%%%%%%%%%%%%%%%%%%%%%%%%%%%%%%%%%%%%%%%%%%%%%%%%%%%%%%%%%%%%%%%%%%

%%%%%%%%%%%%%%%%%%%%%%%%%%%%%%%%%%%%%%%%%%%%%%%%%%%%%%%%%%%%%%%%%%%%%%%%%%%%%%%%%%%%
\newpage

\section{Introduction}\label{sec: Introduction}

%%%%%%%%%%%%%%%%%%%%%%%%%%%%%%%%%%%%%%%%%%%%%%%%%%%%%%%%%%%%%%%%%%%%%%%%%%%%%%%%%%%%

Cryptocurrencies have emerged as an alternative asset class in their own right.
By now, institutional as well as private investors have taken a closer look at this asset class.
Many view digital currencies, particularly the leading cryptocurrency, bitcoin,
as an alternative to gold for providing long--term protection against inflation.
In contrast, a major argument against investing in cryptocurrencies is their high and often erratic volatility
when compared to conventional speculative assets.

Data on bitcoin volatility, in terms of the Bitcoin Volatility Index (BVI), have been available since 2010.${}^1$
The BVI index is based on historical volatility, specifically, the standard deviation of past daily log-returns.
Two versions of this index are provided, based on 30- and 60-day windows of past observations.
The problem with historical volatility is that it is only meaningful if the returns are approximately independent and identically distributed (iid).
In the presence of temporal dependence in the form of volatility clustering,
a phenomenon common to virtually all speculative assets, including cryptocurrencies,
historical volatility fails to adequately capture the current risk of an investment in a speculative asset.
It simply provides an average of past risk.

Implied volatility measures attempt to tackle this shortcoming by capturing the current market view on
future risk as implied by observed transaction prices of option contracts.
Assuming an option-pricing method and given the price at which an option has been traded,
one can back out the expected volatility of the underlying asset that is impled by the observed price.
Several approaches have been developed to price options contracts,
including continuum models such as the \cite{Black_1973} and \cite{Merton_1973} (BSM) formula,
as well as discrete models based on binomial or trinomial trees, Monte Carlo simulation, discrete stochastic volatility,
and finite-differencing methods.

\cite{alexander2021bitcoin} use bitcoin options data from the Deribit exchange to create a term structure
for bitcoin implied volatility indices using the same variance swap fair-value formula (geometric variance swap)
that the Chicago Board Options Exchange (Cboe)
uses to construct the VIX index \citep{VIX_CBOE2019} that reflects the volatility of the S\&P 500 stock index.
Furthermore, Alexander and Imeraj point out that bitcoin prices appear to fluctuate excessively,
so that this approach tends to underestimate the fair value of geometric variance swaps.
In a recent report, \cite{kim2019vcrix} proposed a volatility index based on the cryptocurrency index CRIX.${}^2$
It proxies the next month's mean annualized volatility.
\cite{venter2020garch} used the symmetric GARCH option pricing model to construct various implied volatility indices,
including the CRIX index.
More recently, an alternative volatility index, BitVol \citep{BITVOL_index}, has been introduced.
It reflects the expected 30-day implied volatility derived from tradable bitcoin option prices using the BSM model.  

The BSM model rests upon the assumption that the returns of the underlying asset follow a normal distribution.
There is, however, overwhelming empirical evidence that return distributions exhibit heavy tails and asymmetry.
\cite{osterrieder2016statistics} demonstrated that these properties hold for bitcoin returns.
The omnipresence of extreme return observations indicates that a heavy--tailed distribution should provide a more
realistic description for the behavior of bitcoin returns.
Heavy tails and asymmetry contradict the normality assumption and present a challenge to BSM option pricing.
The Student's $t$--distribution would be an obvious candidate for heavy--tailed data.
However, in the context of option pricing, use of the $t$--distribution in the BSM setting results in a divergent
integral \citep{cassidy2010pricing}. Recently, \cite{naf2019heterogeneous} proposed  a mean–variance heterogeneous tails mixture distribution for modeling financial asset returns.  They showed that the new model captures, along with the obligatory leptokurtosis, different tail behavior among the assets. 

The method of subordination${}^3$ offers an alternative approach to addressing non-normality.
By adopting the concept of intrinsic or operational time rather than calendar time, the method of subordination allows the variance of the normal distribution to change over time and to capture heavy--tailed phenomena.
The subordination approach leads to a generalization of the classical asset pricing model.${}^4$
\cite{shirvani2021multiple} introduced and analyzed various \textit{multiple} subordination processes and demonstrated that the resulting dynamics lead to more realistic asset pricing models.
The idea of multiple subordination is motivated both by the need to better capture the heavy--tailedness that conventional L\'{e}vy--subordinated models fail to explain and the desire to incorporate the views of investors in asset and option pricing models.

In this paper, we develop a bitcoin volatility index based on the Cboe variance swap fair-value formula
but with log--prices following a double subordinated process.
We briefly introduce the idea of double subordination to model the dynamics of L\'{e}vy processes.
We show our model implies a heavy--tailed distribution that provides a better description of bitcoin log--prices
than the Student's $t$ model.

Rather than using Monte Carlo option pricing in non-Gaussian settings, as put forth in \cite{allen2011monte},
we employ the arbitrage theorem and mean--correction martingale measure (MCMM) of \cite{yao2011note} 
to price options when the returns of the underlying assets are doubly subordinated.
Specifically, we price European contingent claims when bitcoin returns are assumed to be driven by a double subordinated L\'{e}vy process.
We determine the equivalent martingale measure to price options using the MCMM approach and demonstrate that our proposed pricing model is arbitrage free.
Based on generated bitcoin option data, we derive implied volatilities using the Cboe approach and the double subordination approach.

The remainder of the paper is organized as follows.
Section \ref{NDIG_section} describes the double subordinated process and applies it to model bitcoin log--prices. 
Section \ref{Bit_optionsection} details the option pricing framework under double subordination and applies it
to model bitcoin vanilla European call and put options.
In Section \ref{BitCoin_Vol} we assess bitcoin volatility using the Cboe approach, which is based upon option pricing,
and our intrinsic time approach based upon spot pricing.
Both methods are compared with historical volatililty.
The paper concludes with a discussion of our results.

\section{ Double Subordination Model for Bitcoin Log-Prices}\label{NDIG_section}
\noindent
The double subordination framework adopted here involves a L\'{e}vy subordinator process.
A stochastic process $\mathbb{X}=(X(t), t \ge 0)$ defined on a stochastic basis $(\Omega,\mathcal{F},\mathbb{F},\mathbb{P})$
is said to be a \textit{L\'{e}vy process} if the following conditions hold.
\begin{itemize}
\item $X\left(0\right)=0,$ $\mathbb{P}$-almost surely.
\item $\mathbb{X}$ is a process with independent increments;
for any partition  $0=t_0<t_1<\dots <t_n<\infty$,
the increments $X\left(t_i\right)-X\left(t_{i-1}\right),\ i=1,2,\dots,n$ are independent.
\item $\mathbb{X}$ is a process with stationary increments;
for any  $0\le s<t$, the increment $X\left(t\right)-X\left(s\right)$ has the same distribution as $X\left(t-s\right) $,
that is, $X\left(t\right)-X\left(s\right)\ {\sim }X\left(t-s\right)$.${}^5$
\item $\mathbb{X}$ is continuous in probability; for every $\varepsilon>0$ and $t\ge 0$, there exists $h_{\varepsilon ,t}>0$,
such that $\mathbb{P}\left(\left|X\left(t+h_{\varepsilon ,t}\right)-X\left(t\right)\right|>\varepsilon \right)<\varepsilon$.
\end{itemize}

In dynamic asset--pricing theory, a risky financial asset is defined by its price dynamics $S_t$, $t\in [0,\tau]$,
where $\tau < \infty$ is the time horizon; that is, $\tau$ is the maturity date of a financial contract.
A L\'{e}vy process $\mathbb{T} = (T(t),\ t \ge 0,\ T(0) = 0)$ with non-decreasing trajectories
(i.e., non-decreasing sample paths) is called a \textit{L\'{e}vy subordinator}.
Since $T\left(0\right)=0$, the trajectories of $\mathbb{T}$ take only non-negative values.
In the BSM option pricing model, the price dynamics of the underlying asset are defined by
\begin{equation}
\label{Black_Eq1}
S_t^{\textrm{(BSM)}} = e^{X_t^{\textrm{(BSM)}}},\quad t\in [0,\tau],
\end{equation}
where the log--process is
\begin{equation}
\label{Black_Eq2}
X_t^{\textrm{(BSM)}} = X_0 + \mu_1 t + \sigma_1 B_{t},\quad \mu_1 \in R,\ \sigma_1 > 0,\ X_0 = \ln(S_0),\ S_0 > 0,
\end{equation}
and $ \mathbb{B} = (B_t,\ t \geq 0)$ is a standard Brownian motion.
To allow for non-normality of asset returns,
\cite{Mandelbrot_1967} and \cite{Clark_1973} suggested the use of a subordinated Brownian motion,
where the price process $S_t^{\textrm{(ss)}}$ and the log--price process are defined as
\begin{equation}
\label{Eq1}
S_t^{\textrm{(ss)}} = e^{X_t^{\textrm{(ss)}}},\quad t\in [0,\tau],
\end{equation}
\begin{equation}
\label{Eq2}
X_t^{\textrm{(ss)}} = X_0+\mu_2 t+\sigma_2 B_{T(t)},\quad \mu_2 \in R,\ \sigma_2 > 0,
\end{equation}
where $\mathbb{T} = (T(t),\ t \geq 0,\ T(0) = 0 )$ is a L\'{e}vy subordinator.

Process \eqref{Eq2} describes the well--known \textit{single} subordinated log--price process.
Various studies have demonstrated that single subordinated log--price models commonly fail to capture the heavy--tailedness observed in financial return data. 
For example, \cite{Lundtofte_2013} and \cite{shirvani2021equity} showed that the normal--inverse Gaussian (NIG) distribution,
which is a single subordinated L\'{e}vy process, fails to explain the equity premium puzzle.
This is partly due to the fact that the tails produced by NIG are not heavy enough.
However, \cite{shirvani2021equity} demonstrated that the high value for the risk aversion coefficient,
which gives rise to the equity premium puzzle,
is compatible with a return process driven by a double subordinator model.
\cite{shirvani2021multiple} defined and investigated the properties of various multiple subordinated log--return processes designed to model leptokurtic asset returns.
They showed that multiple subordinated log--return processes can imply heavier tails than single subordinated models and that they are capable of capturing skewness and kurtosis.
Therefore, a double subordination framework may be a more appropriate candidate for modeling the rather
extreme behavior of bitcoin.

To apply double subordination to modeling the bitcoin price process, let $S_t$ denote the price process with the dynamics
\begin{equation}
\label{DSS}
S_t = e^{X_t},\quad t\in [0,\tau],
\end{equation}
\begin{equation}
\label{DSX}
X_t=X_0+\mu_3 t+\gamma U(t) + \rho T(U(t)) +\sigma_3 B_{T(U\left(t\right))},\quad t \ge 0,\  \mu_3 \in R,\ \sigma_3 > 0,\ X_0 = \ln(S_0),\ S_0 > 0,
\end{equation}
where the triplet  members $\{B_s, T(s), U(s)\}, s \ge 0$,
are independent processes generating the stochastic basis
$(\Omega,\mathcal{F},\ \mathbb{F}=\left({\mathcal{F}}_t,t\ge 0\right), \mathbb{P})$
representing the real world.
Here, $\{ B_s,\ s \ge 0 \} $ is a standard Brownian motion,
while $\{ T(s),\ s \ge 0,\ T(0)=0 \} $ and  $\{ U(s),\ s \ge 0,\ U(0)=0\} $  are L\'{e}vy subordinators.
$B_t$, $T(t)$ and $U(t)$ are ${\mathcal{F}}_t$-adapted processes,
whose trajectories are right-continuous with left limits.
\cite{shirvani2021multiple} referred to $T(U(t)),t\ge 0$ as the \textit{double subordinator process};
hence, the process modeled by \eqref{DSX} is a \textit{double subordinated log--price process}.

Consider the case where the subordinators $T(t)$ and $U(t)$ are inverse Gaussian (IG) L\'{e}vy processes;
that is, $T(1) \sim IG(\lambda_T, \mu_T)$ having the probability density function (pdf),
\begin{equation}
\label{IG_dis}
f_{T(1)}(x)
= \sqrt{\frac{{\lambda }_T}{2\pi x^3}}{\mathrm{exp} \left(-\frac{{\lambda }_T{\left(x-{\mu }_T\right)}^2}{2{\mu }^2_Tx}\right)\ },
\quad x \ge 0,\ \lambda _T > 0,\ \mu _T > 0.
\end{equation}
Similarly $U\left(1\right)\sim IG\left({\lambda }_U, {\mu }_U\right)$. 
In this case, \cite{shirvani2021multiple} referred to $T(U(t))$ as the \textit{double inverse Gaussian subordinator}
and to $X_t$ as a \textit{normal double inverse Gaussian (NDIG) log--price process}.
The characteristic function (chf) of $X_1$ is given by

\begin{equation}
\label{chf_double_IG_Log_price}
\begin{array}{cc}
\varphi_{X_1}(v) = \textrm{E} \left[ e^{i v X_1 } \right]=\\
\exp \left\{
i v \mu_3 + \frac{\lambda_U}{\mu_U}
\left[ 1-\sqrt{
1-\frac{2\mu_U^2}{\lambda_U}
\left( \frac{\lambda_T}{\mu_T}
\left( 1-\sqrt{1-\frac{\mu_T^2}{\lambda_T}(2iv\rho -\sigma_3^2 v^2) }  \right)+iv\lambda
\right)
}
\ \right]
\right\},
\end{array}
\end{equation}
with $v \in \mathbb{R}$.
The moment generating function (MGF) of $X_1$ is
$M_{X_1}(w) = \textrm{E} \left[ e^{w X_1 } \right]$, $w \in \mathbb{R}$.

The NDIG model \ref{DSX} and \ref{chf_double_IG_Log_price} has eight parameters, namely $\mu_3$, $\sigma_3$, $\gamma$, $\rho$, $\mu_{\tau}$, $\lambda_{\tau}$, $\mu_T$ and $\lambda_T$, which can make fitting the model to data a challenging task. However, it is worth noting that that only six of these parameters are identifiable within the model.\footnote{See \cite{Lindquist2022}.} To set the remaining two parameters, we can consider the expectation:

\begin{equation}
\mathbb{E} \left[ X_1 \right] = \mu_3 + \gamma \mathbb{E} \left[ U(1) \right] + \rho \mathbb{E} \left[ X(v) \right] \mathbb{E} \left[ T(U(1)) \right]
\end{equation}

As the processes $U$ and $T$ are independent inverse Gaussian (IG) processes, we can uniquely identify $\gamma$ and $\rho$ by requiring:

\begin{equation}
\label{Exp_1}
\mathbb{E} \left[ U(1) \right] = \mu_U = 1 \quad \text{and} \quad \mathbb{E} \left[ T(U(1)) \right] = \mu_U \mu_T = 1 \quad \rightarrow \quad \mu_T = 1.
\end{equation}

By utilizing equation \ref{Exp_1}, the set of model parameters that can be identified becomes $\mu_3$, $\sigma_3$, $\gamma$, $\rho$, $\lambda_{T}$, and $\lambda_U$.

The central moments of the NDIG can serve as a source of data for fitting these six parameters. The moment generating function (MGF) $M_{X_1}(w)$ for $X_1$, which generates the moments of its probability distribution, can be obtained by evaluating $\mathbb{E} \left[ e^{iwX_1} \right]$ for $w \in \mathbb{R}$. This can be obtained from equation \ref{chf_double_IG_Log_price} by setting $w = iv$. The MGF is written in terms of the cumulant generating function $K(w)$:

\begin{equation}
\label{cum_gen_fun_log_price}
M_{X_1}(w) = \mathbb{E} \left[ e^{iwX_1} \right] = \exp \left( K_{X_1}(w) \right).
\end{equation}

\noindent Using the identity 
\begin{equation}
\label{power_t}
\varphi_{X_t}(v) = \mathbb{E} \left[ e^{ivX_t} \right] = ( \mathbb{E} \left[ e^{ivX_1} \right] )^t =  \left[ \varphi(X_1) \right]^t,
\end{equation}
we can express $M_{X_t}(w)$ as $M_{X_t}(w) = \exp \left( K_{X_1}(w) t\right)$, where $K_{X_1}(w)$ can be written using equation \ref{cum_gen_fun_log_price} as:

\begin{equation}
\label{K_X_new}
\begin{aligned}
K_{X_1}(w) &= \mu_3 w + \lambda_U \left( 1 - \sqrt{1 - g(w)} \right), \\
g(w) &= 1 - 2 \frac{\lambda_T}{\lambda_U}\left(1 - \sqrt{ h(w)}\right) - 2 \frac{\gamma }{\lambda_U}w, \\
h(w) &= 1 - \frac{2\rho w}{\lambda_T}-\frac{\sigma_3^2 w^2}{ \lambda_T}.
\end{aligned}
\end{equation}

\noindent Now, by finding the first four central derivatives of $K_{X_1}(w)$, we can have the first four centered moments of $X_1$.

\begin{equation}
\label{eq_moments}
\begin{aligned}
E[X] &= \mu_3 + s, \\
Var[X] &= \sigma + \frac{s^2}{\lambda_U}, \\
Skew[X] &= \frac{3\left(\sigma c+ \frac{s^3}{\lambda_U^2}\right)}{\left(\sigma+\frac{s^2}{\lambda_U}\right)^{3/2}}, \\
Kurt[X] &= 3 \frac{ [\sigma^2\left(1/\lambda_T +1/\lambda_U \right) + 2\sigma \left( c^2+(\rho/\lambda_T)^2+2(s/\lambda_U)^2 \right) + 5  s^4/\lambda_U^3]}{\left(\sigma + \frac{s^2}{\lambda_U}\right)^2} ,
\end{aligned}
\end{equation}
where $s = \gamma+\rho$, $\sigma = \rho^2/\lambda_T + \sigma_3^2$, and $c = {\rho/\lambda_T}+{s/\lambda_U}$.
Equation \ref{eq_moments} presents four criteria for fitting the six model parameters. The subsequent section on numerical computation will discuss additional requirements for parameter estimation.

\section{ Bitcoin Option Pricing under Double Subordination}\label{Bit_optionsection}
\noindent
To price European contingent claims, we assume that $X_t$ follows a NDIG log--price process.
We need to derive an equivalent martingale measure (EMM)  $\mathbb{Q}$ of $\mathbb{P}$ on
$(\Omega,\mathcal{F},\ \mathbb{F}=\left({\mathcal{F}}_t,t\ge 0\right), \mathbb{P})$,
such that the discounted price process, $e^{-rt} S_t$ with $r\geq 0$ denoting the riskless rate, is a martingale.${}^7$
To do so, we use the MCMM, since the proposed pricing model is arbitrage-free under MCMM,
and estimate the parameters specifying the process.
We then add the appropriate drift term to the process, such that the discounted price process becomes a martingale.

\cite{yao2011note} constructed a martingale measure using MCMM for the geometric L\'{e}vy process model
and showed that this measure is an EMM if there is a continuous Gaussian part in the L\'{e}vy process.
In the case that $X_t$ is a pure jump L\'{e}vy process,
they pointed out that this measure cannot be equivalent to a physical probability.
However, pricing European options under this measure is still arbitrage free.

Let $\mathcal{C}$ be a European call option having an underlying risky asset $\mathcal{S}$
with a price and log--price process as in \eqref{DSS} and \eqref{DSX}, respectively.
Let $\mathcal{B}$ be a riskless asset with price $b_t=e^{rt},t\ge 0$, where $r\ge 0$ is the riskless rate.
Then, the price of $\mathcal{C}$ is
\begin{equation}
\label{call_price}
C (S_0,r,K,\tau )=e^{-r\tau} \ \textrm{E}_{\mathbb{Q}} \left[ \mathrm{max}  (S^{(\mathbb{Q} )}_\tau - K,0 ) \right],
\end{equation}
where  $\tau>0$ denotes the maturity, $K>0$ is the strike price, and $S_t^{ (\mathbb{Q} )}$ is the price dynamic of $\mathcal{S}$ on $\mathbb{Q}$ (an equivalent martingale measure of $\mathbb{P}$).
Using MCMM to find the EMM,
\begin{equation}
\label{Eq3}
S^{(\mathbb{Q})}_t=\frac{b_t S_t}{M_{X_t} (1)}=S_0 e^{\left( r - K_{X_1}(1) \right) t + X_t},\quad t \in [0,\tau],
\end{equation}
describes the dynamics of $\mathcal{S}$ on $\mathbb{Q}$, where $M_{ X_t } (v)$ is the MGF of $X_t$ and 
$K_{ X_t}(w) = \ln M_{X_t}(w)$ is the cumulant--generating function of $X_t$.
The chf of the log--price, ${\ln S^{ (\mathbb{Q} )}_t}$, is
\begin{equation}
\label{chf_Log_price}
{\varphi }_{\ln S^{ (\mathbb{Q} )}_t} (v )=S^{iv}_0{\mathrm{exp}  \{ [iv (r-K_{X_1}(1))+{\psi }_{X_1} (v) ]t\}},\quad t \in [0,\tau],
\end{equation}
where ${\psi }_{ X_t} (v)$, $v \in \mathbb{R}$ is the characteristic exponent of $X_t$.

\cite{Carr_1998} showed how to use the fast Fourier transform (FFT) to value options in the case in which the chf of the log-price of the
underlying asset is known analytically.
They consider the modified option price $c_a (S_0,r,k,\tau) = e^{a k} C (S_0,r,k,\tau )$, where $k =\ln(K)$, which, for
a range of values $a>0$, guarantees that $c_a (S_0,r,k,\tau)$ is square integrable over $k \in (-\infty,\infty)$.
The access point for applying the FFT is their derived relationship
\begin{equation}
\label{Carr_call}
C (S_0,r,k,\tau ) = \frac{e^{-r\tau-ak}}{\pi }
\int^{\infty }_0{e^{-ivk}} \frac{{\varphi }_{\ln S^{(\mathbb{Q})}_\tau} (v - i (a+1))} {a^2 + a - v^2 + i (2a + 2)v} dv.
\end{equation}
Numerical solution of this integral involves two fundamental concerns, an ``optimum'' value for $a$
and control over the error produced by truncating the integral \eqref{Carr_call} over a finite domain $[0,v_{\textrm{max}}]$.
Addressing the first of these two concerns, Carr and Madan note that a positive value for $a$ guarantees square-integrability of
$c_a (S_0,r,K,\tau)$ over the negative $k$ axis but aggravates square-integrability over the positive $k$ axis.
They derive a sufficient condition, $\textrm{E}_{\mathbb{Q}} \left[ \left( S^{(\mathbb{Q})}_\tau \right)^{a+1} \right] < \infty $
which, combined with the analytic expression for the chf, can be used to determine an upper bound on $a$.
Addressing the second of these concerns, they develop a bound on the error involved in truncating the integration range of \eqref{Carr_call},
\begin{equation*}
v_{\textrm{max}} > \frac{\exp(-ak)}{\pi} \frac{\sqrt{A}}{\epsilon},
\end{equation*}
where $\epsilon$ is a desired level of truncation error and $\sqrt{A}/v^2$ is an upper bound on the magnitude of the integrand in \eqref{Carr_call}.

Constructing the implied volatility surface for call prices requires estimating \eqref{Carr_call} over a discrete mesh of $(k,T)$ values.
Put options can then be valued assuming that put–call parity holds.
Implementing a FFT requires numerical discretization of \eqref{Carr_call} into the form
\begin{equation}
\label{FFT_sum}
\hat{Z}_p = \sum_{j=1}^{M} \exp\left[ -i \frac{2\pi}{M} (j-1) (p-1) \right] Z_j, \quad k=1,2,\dots M,
\end{equation}
which the FFT can solve in $O(M \ln_2 (M))$ operations.
Writing \eqref{Carr_call} as
\begin{equation}
C(\cdot,\tau,k) = \frac{e^{-r\tau-ak}}{\pi }\Psi(k), \quad \textrm{where } \Psi(k) = \int_{0}^{\infty} exp (-ivk)h(v)dv,
\label{FFT_G}
\end{equation}
discretization of $\Psi(k)$ over a finite range $[0,v_\textrm{max}]$ using the left hand rectangle rule${}^8$ gives
\begin{equation}
\label{Psi_rect}
\Psi(k)  = \sum_{j=1}^N \exp(-i v_j k) h(v_j)\Delta v,
\end{equation}
where $v_j = (j-1)\Delta v$, $j = 1, ..., N$, $v_1 = 0$, $v_N = v_\textrm{max}-\Delta v$.
For each fixed value of $\tau$, we discretize $k=\ln (K)$ over a range $[-\bar{k}, \bar{k}]$ with $N$ equally spaced points,
$k_p = -\bar{k} +(p-1)\Delta k$, $p = 1, ..., N$, $\Delta k = 2\bar{k}/N$.
On this grid, \eqref{Psi_rect} becomes
\begin{equation}
\label{Psi_relab2}
\Psi(k_p)  = \sum_{j=1}^{N} \exp[-i (j-1) \Delta v (p-1)\Delta k] \exp(i \bar{k} v_j) h(v_j)\Delta v,
\quad p = 1, ..., N,
\end{equation}
which is identical to \eqref{FFT_sum} with the identifications
$\hat{Z}_p=\Psi(k_p)$, $Z_j = \exp(i \bar{k} v_j) h(v{j})\Delta v$,
$M = N$, and $2\pi/M = \Delta v \Delta k$.
This last identification gives the familiar FFT tradeoff between the span covered in the ``space'' domain and in the
``frequency'' domain,
\begin{equation}
\label{vmaz}
v_{\textrm{max}} \bar{k} = \pi N.
\end{equation}
In our computations in section~\ref{NDIG_option}, we used $N=2^{10}=1024$.

\cite{Carr_1998} introduced the parameter $a$ to ensure that the call pricing function \ref{call_price} is square integrable as $K \rightarrow 0$ (i.e., as $k =lnK\rightarrow -\infty$). They note that a sufficient condition for square integrability is provided by the requirement that $\phi_{ln(S_t^{\mathbb{Q}})}(-i(1+a)) \leq \infty$.

\noindent From \ref{chf_Log_price}, 
\begin{equation}
\label{chf_a}
{\varphi }_{\ln S^{ (\mathbb{Q} )}_t} (-i(1+a) )=S^{1+a}_t 
\mathrm{exp} \{ (1+a) \left(  r_f - K_{X}(1)   \right) T \} 
\{ \mathrm{exp}  \left(  K_{X} (1+a)  \right)  \}^T
\leq \infty.
\end{equation}
From \ref{chf_double_IG_Log_price} and \ref{cum_gen_fun_log_price}, note that $\psi(iw) \leq K\psi(w)$ for $w \in \mathbb{R}$. Hence, \ref{chf_a} and $\phi_{ln(S_t^{\mathbb{Q}})}(-i(1+a)) \leq \infty$ can be combined as the requirement 

\begin{equation}
\label{chf_a_condition}
\mathrm{exp} \{ (1+a) \left(  r_f - K_{X}(1)   \right) T \} 
\{ \mathrm{exp}  \left(  K_{X} (1+a)  \right)  \}^T
\leq \infty.
\end{equation}

\noindent To ensure that the cumulant generating function remains real-valued, requirement \ref{chf_a_condition} can be reduced to positive argument requirements for the square root evaluations in \ref{K_X_new} for $K_{X_1} \in [1, a]$, where $a \in (0, \infty)$. From \ref{cum_gen_fun_log_price}, under the assumption that $\gamma = 0$, this can be further reduced to the following requirements:

\begin{equation}
\label{a_conditon_2}
\begin{aligned}
h(w)= 1 -\frac{\sigma_3^2 w^2}{ \lambda_T} - \frac{2\rho w}{\lambda_T} \geq 0 \,\,\,\,\, \text{and} \,\,\,\,\,
g(w)= 1- 2 \frac{\lambda_T}{\lambda_U}\left(1 - \sqrt{h(w)}\right) \geq 0 . 
\end{aligned}
\end{equation}
Solving the latter equation for $h(w)$ yields
\begin{equation}
\label{a_conditon_23}
h(w) \geq 1- \frac{\lambda_U}{2\lambda T} \stackrel{\mathrm{def}}{=}  d^2.
\end{equation}
Combining \ref{a_conditon_23} with the first equation in \ref{a_conditon_2} gives
\begin{equation}
\label{a_conditon_3}
1 -\frac{\sigma_3^2 w^2}{ \lambda_T} - 2\frac{\rho w}{ \lambda_T} \geq d^2.
\end{equation}
The quadratic equation has two roots, and Equation \ref{a_conditon_3} is satisfied for $w=1+a$ when 
\begin{equation}
\label{a_conditon_4}
-\frac{\rho}{\sigma_3^2} -
\frac{ \sqrt{ \rho^2+ \lambda_t \sigma_3^2 (1-d^2)  }     }
{\sigma_3^2}
\leq
1+a
\leq
-\frac{\rho}{\sigma_3^2} +
\frac{ \sqrt{ \rho^2+ \lambda_t \sigma_3^2 (1-d^2)  }     }
{\sigma_3^2}
\end{equation}
Since $a$ is a non-zero value, it follows that
\begin{equation}
\label{a_conditon_5}
a \leq a_{max} = 
1-\frac{\rho}{\sigma_3^2} +
\frac{ \sqrt{ \rho^2+ \lambda_t \sigma_3^2 (1-d^2)  }     }
{\sigma_3^2}
\end{equation}
By utilizing \ref{a_conditon_3}, equation \ref{a_conditon_5} can be rewritten as

\begin{equation}
\label{a_conditon_final}
a_{max} = 
\frac{1}
{\sigma_3^2} 
\sqrt{  \rho^2 + \lambda_U \left( 1 - \frac{\lambda_U}{4 \lambda_T}  \right) \sigma_3^2   }
- \frac{\rho}{\sigma_3^2} -1.
\end{equation}

\section{Numerical computation}
\begin{figure}[h]
\begin{center}
\begin{subfigure}[b]{0.49\textwidth}
\includegraphics[width=\textwidth]{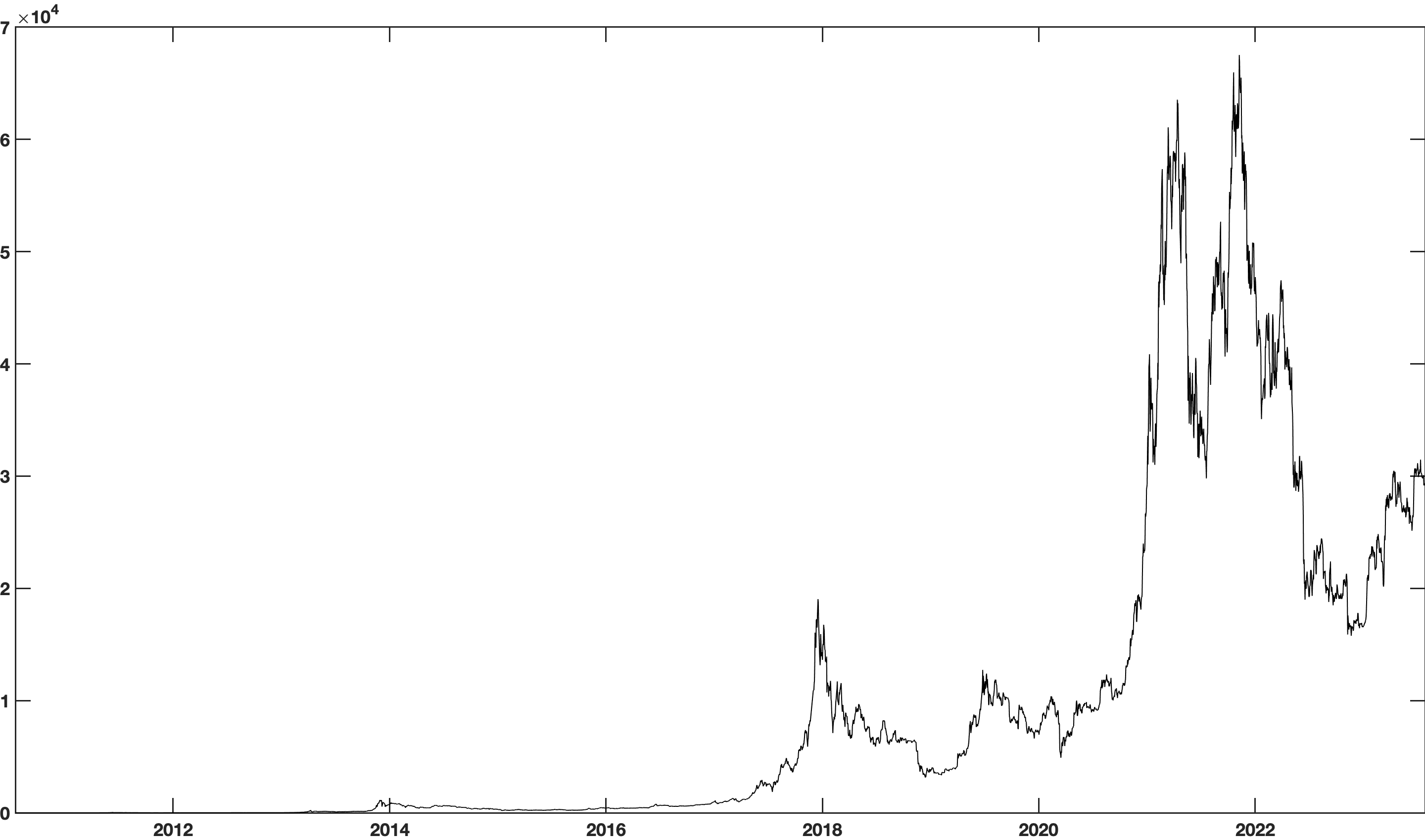}
\label{fig_1_a}
\end{subfigure}
\begin{subfigure}[b]{0.49\textwidth}
\includegraphics[width=\textwidth]{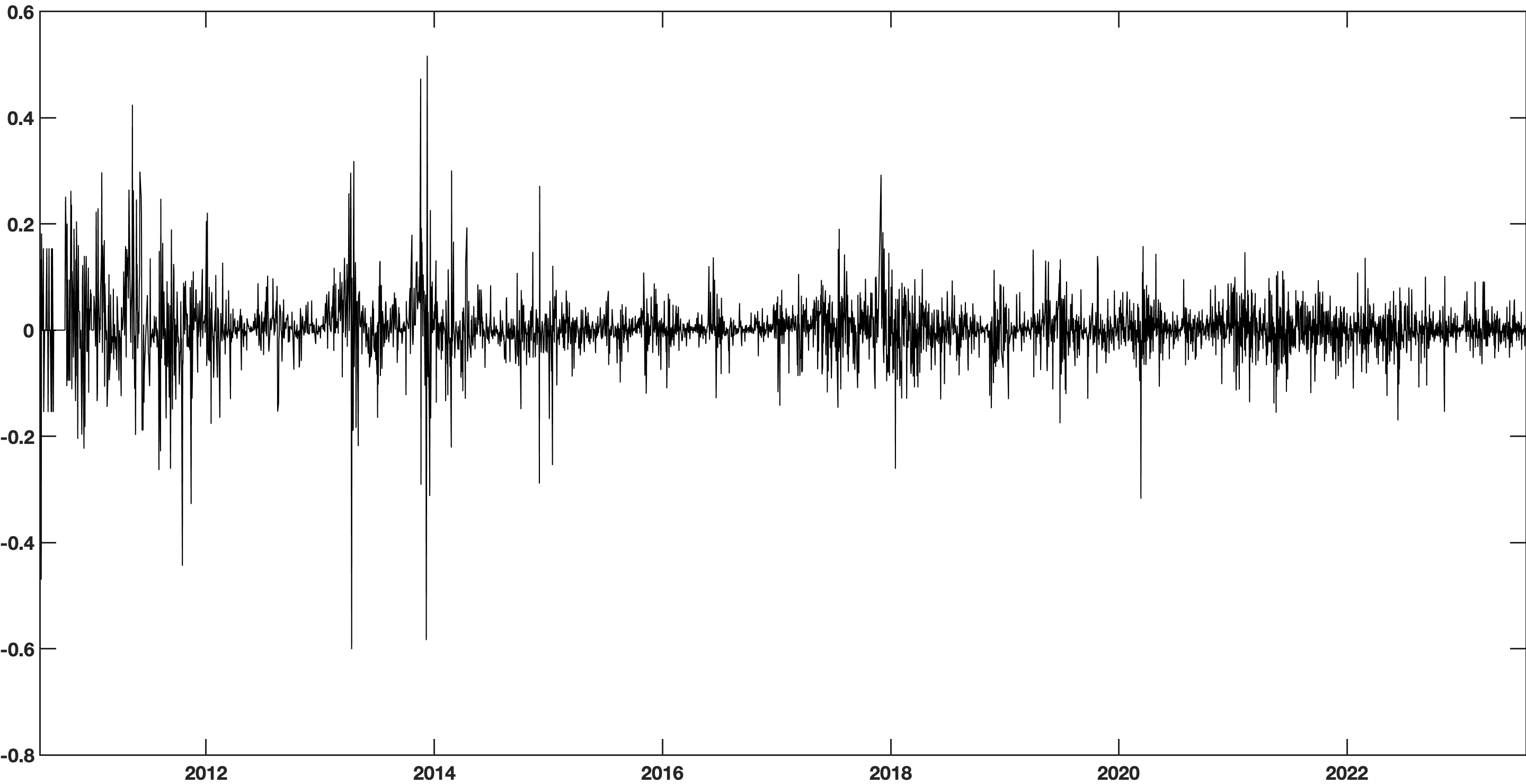}
\label{fig_1_b}
\end{subfigure}
\captionsetup{labelfont=bf,labelsep=quad}
\caption{Daily bitcoin data from July 19, 2010 to  July 28, 2023.}
\label{Fig_1}
\end{center}
\end{figure}

In this section, we illustrate the double subordinated method of sections \ref{NDIG_section} and \ref{Bit_optionsection} and fit the NDIG model to the log-return Bitcoin time series. 
 Figure~\ref{Fig_1} shows the daily bitcoin price and log-return, $r_t$
time series covering the period from July 19, 2010 to July 28, 2023.
The return series consists of 4,026 observations, to which we fit the NDIG model specified by
\eqref{DSS}--\eqref{chf_double_IG_Log_price}. 
The NDIG model is characterized by six parameters, $\theta = (\mu_3, \lambda_U, \mu_U, \lambda_T, \mu_T, \sigma_3)$.
To further simplify the parameter set $\mu_3$, $\sigma_3$, $\gamma$, $\rho$, $\lambda_U$, and $\lambda_T$, we assume that the subordinator $U(t)$ is used to model the intrinsic time of the return process, while the subordinator $T(t)$ is used to model the return skewness and heavy-tailed behavior. It is reasonable in this model to demand that there be no $\gamma U(t)$ term in \ref{DSX}, i.e., $\gamma=0$. With this requirement, the first four moments of $X$ become:

\begin{equation}
\label{eq_moments}
\begin{aligned}
E[X] &= \mu_3 + \rho, \\
Var[X] &= \sigma_3^2 + d\rho^2, \\
Skew[X] &= \frac{3d\rho\left(\sigma_3^2 + \rho^2\right)}{\left(\sigma_3^2+d\rho^2\right)^{3/2}}, \\
Kurt[X] &=  \frac{3\sigma^2 d + 6 \rho^2 \sigma \left(
d^2+1/\lambda_T^2 +2/\lambda_U^2
\right) + \frac{15\rho^4}{\lambda_U^3}
}
{\left(\sigma_3^2+d\rho^2\right)^2} ,
\end{aligned}
\end{equation}
where $\sigma =\rho^2/\lambda_T + \sigma_3^2$, $d = 1/{\lambda_T}+1/{\lambda_U}$. 

As there is no analytical expression for the probability density function, we employ the method of moments and the empirical chf to estimate the parameters of the model. We follow \cite{Paulson_1975} and \cite{Yu_2003}
and use the fact that the pdf is the Fourier transform of the chf.
So, the five parameters $\mu_3$, $\sigma_3$, $\rho$, $\lambda_U$, and $\lambda_T$ can be estimated via minimization

\begin{mini}|s|
{\mu_3,\sigma_3, \rho, \lambda_T,\lambda_U}{   \{   
 \left(  \Delta M_1  \right)^2+ \left(  \Delta M_2  \right)^2+ \left(  \Delta M_3  \right)^2+ \left(  \Delta M_4  \right)^2+        \left(  \Delta CF  \right)^2      \}   }
{\label{eq:Example1}}
{}
\addConstraint{\left(  \Delta M_1  \right)^2   }{=1- \frac{\mathrm{E}[X_t]}{\mathrm{E}[r_t]} }
\addConstraint{\left(  \Delta M_2  \right)^2}{=  1- \frac{\mathrm{Var}[X_t]}{\mathrm{Var}[r_t]}}
\addConstraint{ \left(  \Delta M_3  \right)^2 }{= 1- \frac{\mathrm{Skew}[X_t]}{\mathrm{Skew}[r_t]} }
\addConstraint{ \left(  \Delta M_4  \right)^2 }{= 1- \frac{\mathrm{Kurt}[X_t]}{\mathrm{Kurt}[r_t]}}
\addConstraint{ \left(  \Delta CF  \right)^2   }{= \int_{-\infty}^{\infty}\left( \frac{1}{n}\sum_{j=1}^{n} e^{iv x_j} -\varphi_{X_t} \left( v,\theta\right)\right)^2 dv.}
\end{mini}
where $r_t$  denotes the observed  daily bitcoin return time series. The inclusion of the term $\Delta CF$ relies on the one-to-one correspondence between the cumulative distribution function and the characteristic function (since the probability density function is the inverse Fourier transform of the characteristic function). The integral for $\Delta CF$ can be estimated using the method described by \cite{Yu_2003}.

The parameters are estimated by the
method of moments and the empirical chf fitting, which is a method as efficient as likelihood--based methods
\citep{Yu_2003}. 

The resulting parameter estimates are reported in Table~\ref{NDIG_Par}.

\begin{table}[h]
\centering
\captionsetup{labelfont=bf,labelsep=quad}
\caption{NDIG parameter estimates for the bitcoin log returns.}
\label{NDIG_Par}
\begin{tabular}{@{}ccccccc@{}}
\toprule
$\mu_3$     & $\lambda_U$     & $\mu_U$  & $\lambda_T$   & $\mu_T $    & $\sigma_3$  & $\rho$ \\ \midrule
0.004      & 0.145          & 1.00   & 9.9293        & 1.00     & 0.0551      & -0.0008\\ \bottomrule
\end{tabular}
\end{table}

\begin{figure}[hbt]
\begin{center}
\begin{subfigure}[b]{0.49\textwidth}
\includegraphics[width=\textwidth]{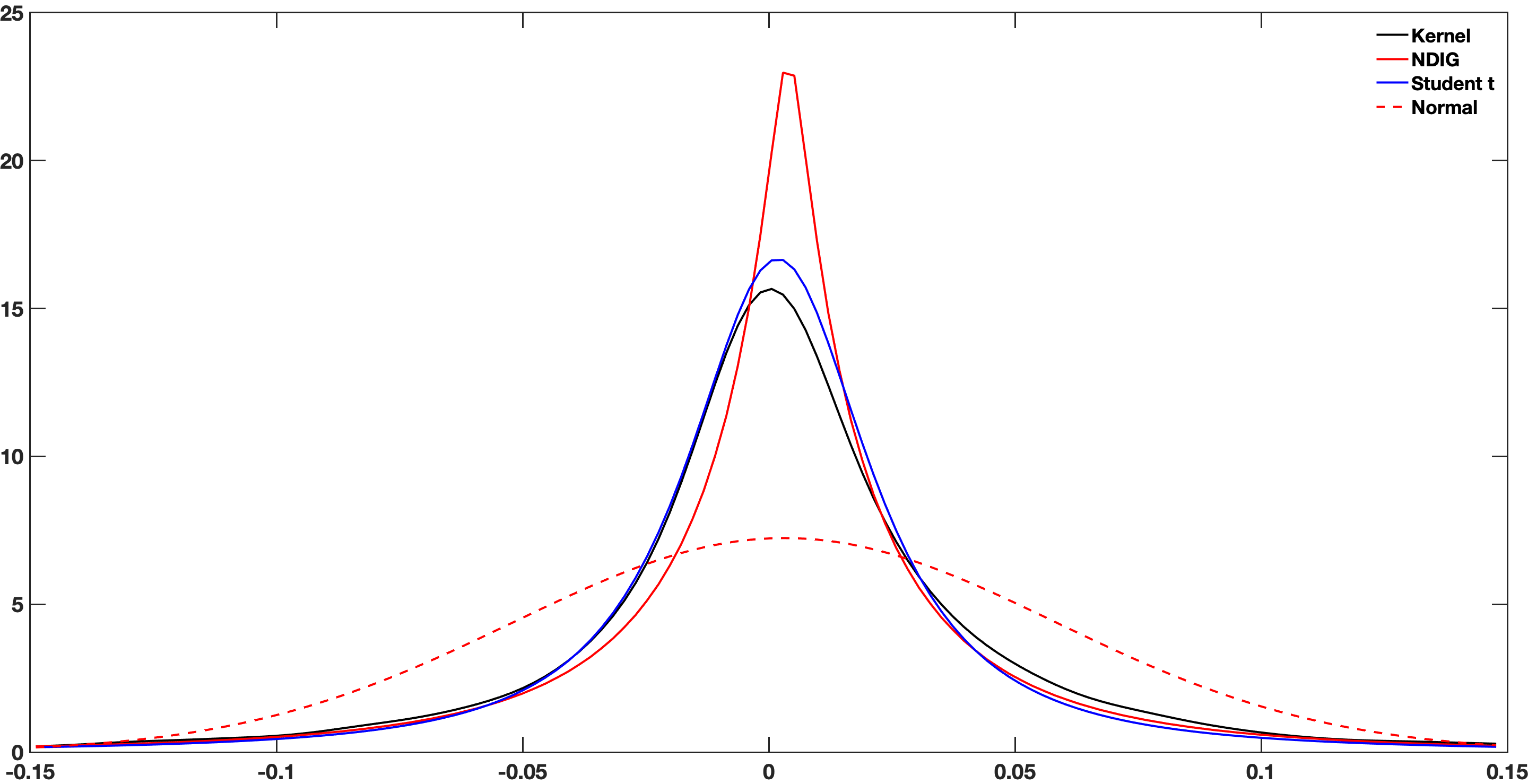}
\caption{}
\label{fig_2_a}
\end{subfigure}
\begin{subfigure}[b]{0.49\textwidth}
\includegraphics[width=\textwidth]{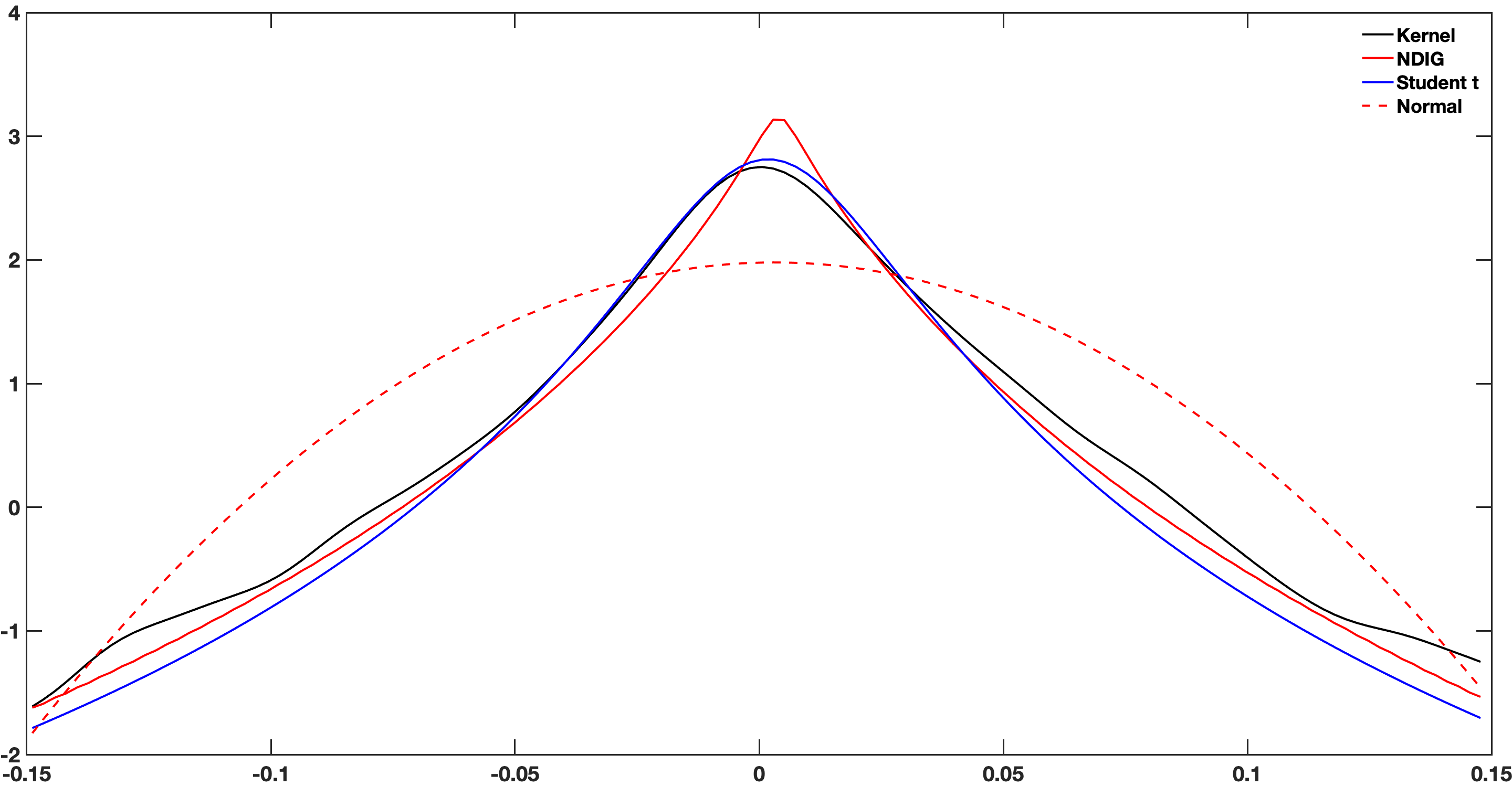}
\caption{}
\label{fig_2_b}
\end{subfigure}
\captionsetup{labelfont=bf,labelsep=quad}
\caption{(a) Comparison of normal, Student's $t$, and NDIG distribution fits to the kernal density of the bitcoin return.
(b) Same plot with the y-axis now in natural log scale.
}
\label{Ksdensity_NDIG}
\end{center}
\end{figure}

For comparison, we also fit the normal and the Student's $t$ distributions to the Bitcoin daily log-return values
and estimated the distribution parameters using maximum likelihood. 
The empirical density, the fitted NDIG model, and the best-fit Student's $t$ ($df=1.60$),
and normal ($\hat{\mu} = 0.0032$, $\hat{\sigma} = 0.0551$) distribution densities are compared in Figure~\ref{Ksdensity_NDIG}. 
The results confirm, as was noted in the introduction, that the  normal distribution is not well suited for modeling
asset returns.
The graphs indicate that both the Student's $t$ and NDIG models match the tails of empirical density rather well,
while NDIG matches the skewness in the empirical distribution more evenly than does Student's $t$. 
The low estimate of $df = 1.60$ for the Student's $t$ distribution implies that even second moments do not exist. 
Thus, the fat tails of the Student's $t$ distribution would cause divergence of the BSM integral needed to evaluate
price options.

\subsection{NDIG Model for European Call Option Pricing}\label{NDIG_option}

In this section, the NDIG model is used to price a European call option $\mathcal{C}$ with the underlying risky asset $\mathcal{S}$ being the return of Bitcoin. The dynamics of $\mathcal{S}$ under $\mathbb{Q}$ is given by equation \ref{Eq3}, and the characteristic function (CF) of the log-price by equation \ref{chf_Log_price}. Option prices for $\mathcal{C}$ are determined by evaluating equation \ref{Carr_call} using the Fast Fourier Transform (FFT) for different values of the strike price $K$ and maturity time $T$. Put option prices are computed using the put-call parity.

We apply the  NDIG L\'{e}vy model to the pricing of European plain vanilla bitcoin options.
Let $\mathcal{C}$ be a European call option where the underlying risky asset, $\mathcal{S}$,
follows the log--price process given in  \eqref{DSX}.
The dynamics of $\mathcal{S}$ on $\mathbb{Q}$ are given by \eqref{Eq3}
and the chf of the log--price by \eqref{chf_Log_price}.
We evaluate the integral \eqref{Carr_call} using FFT for a range of strike levels and maturity horizons with the parameter estimates reported in Table \ref{NDIG_Par}. When using the global parameter values provided in Table \ref{NDIG_Par}, the upper bound of $a$ in \ref{eq_moments}  is $ \lesssim 21.93$.

To ensure that option prices are within acceptable limits, a more stringent determination of the value of $\sigma \in (0,\infty)$ is required. Specifically, for a European call option, the upper and lower bounds on the price are given by $\max(S-K e^{-r_f T,}, 0) \leq C(S_t,T,K) \leq S_T $.\footnote{See, e.g., \cite{Hull_2018}.}

As noted above, an upper bound value for $a$ and hence choice of a suitable value in performing the integrations via FFT
is very dependent on the data set (the price process).
When $a > 1$ the call option surface was unstable.
We follow \cite{Schoutens_2003} and set $a=0.15$.

\begin{figure}[h]
\begin{center}
\begin{subfigure}[b]{0.49\textwidth}
\includegraphics[width=\textwidth]{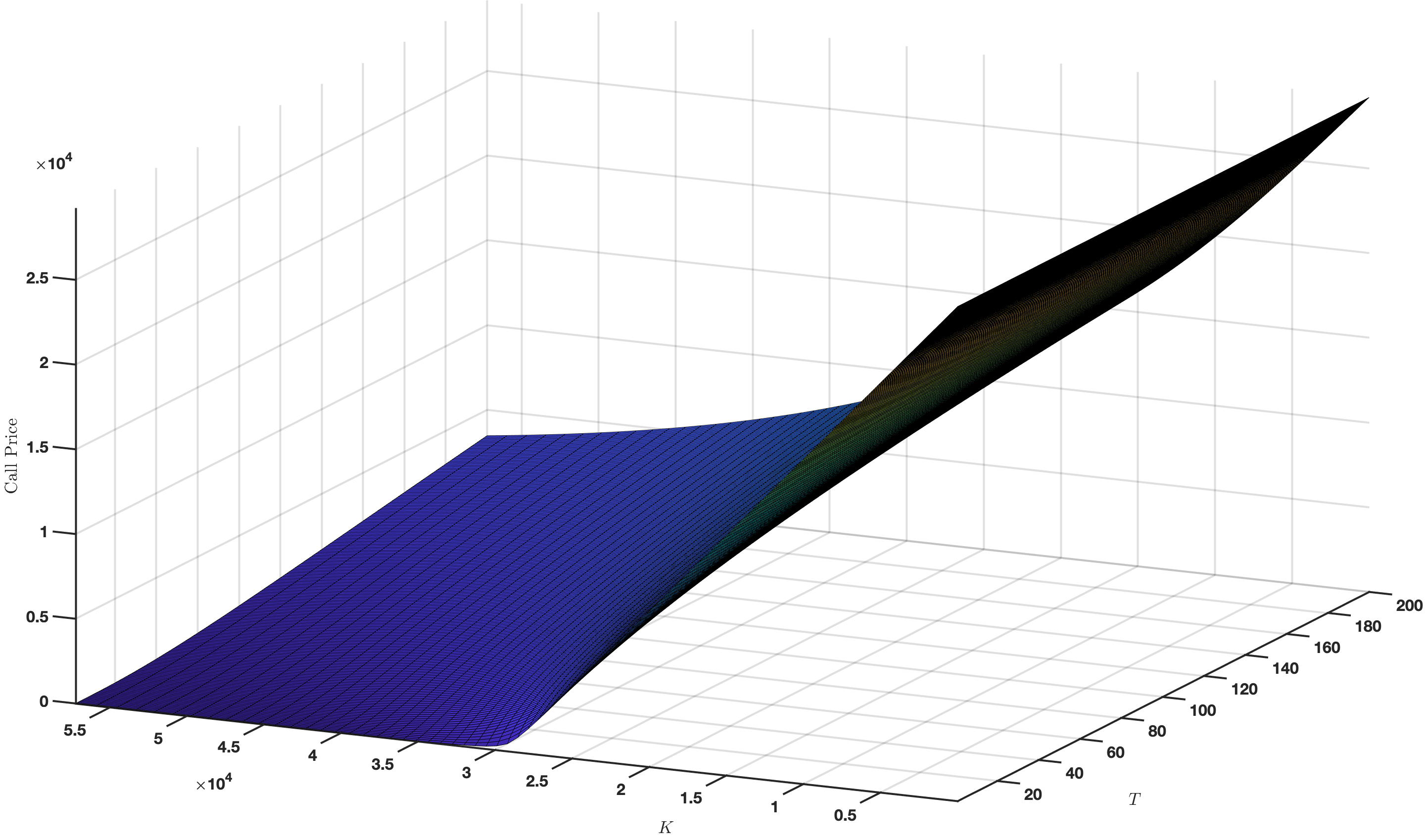}
\caption{}
\label{Call_price_NDIG}
\end{subfigure}
\begin{subfigure}[b]{0.49\textwidth}
\includegraphics[width=\textwidth]{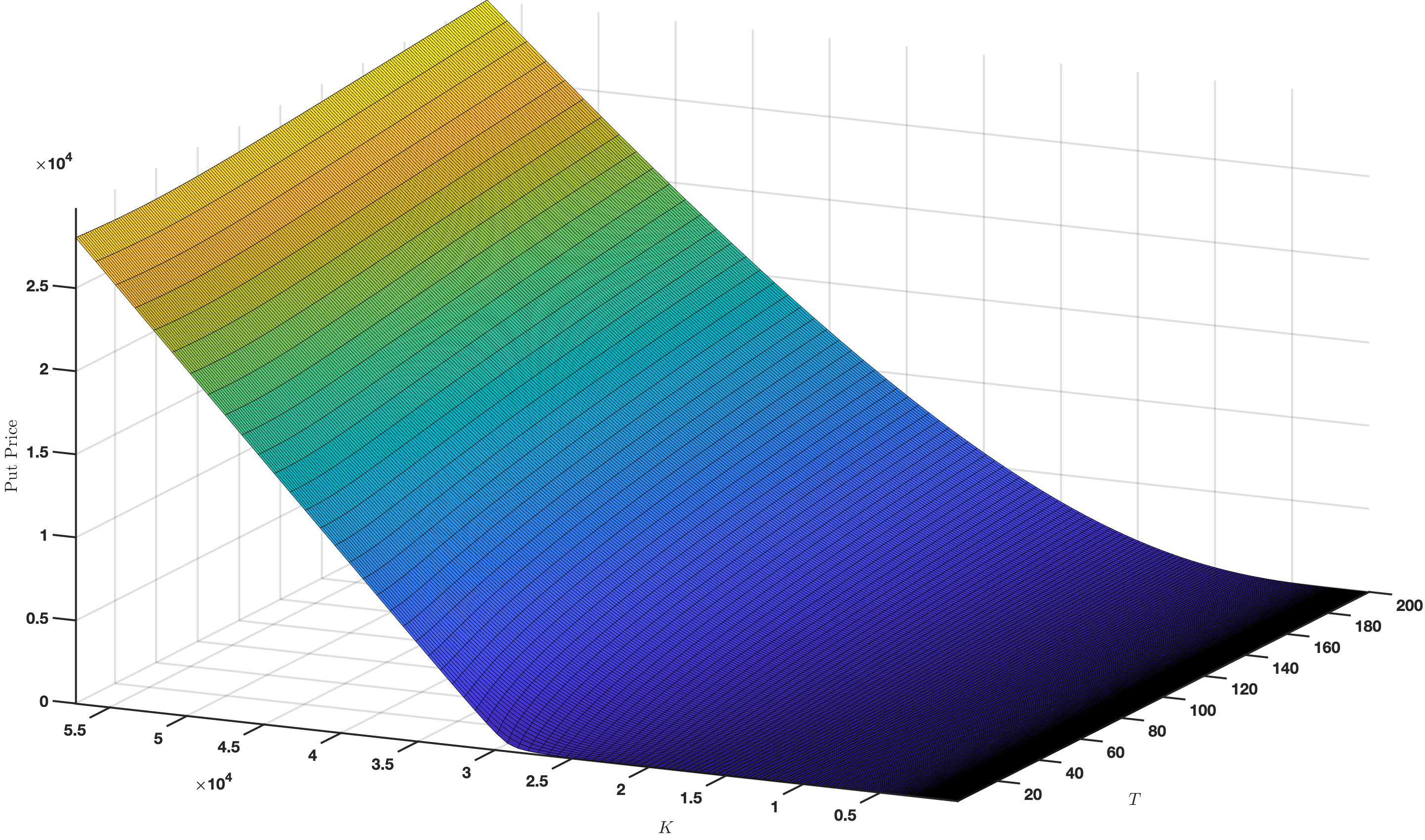}
\caption{}
\label{Put_price_NDIG}
\end{subfigure}
\par\medskip
\begin{subfigure}[b]{0.49\textwidth}
\includegraphics[width=\textwidth]{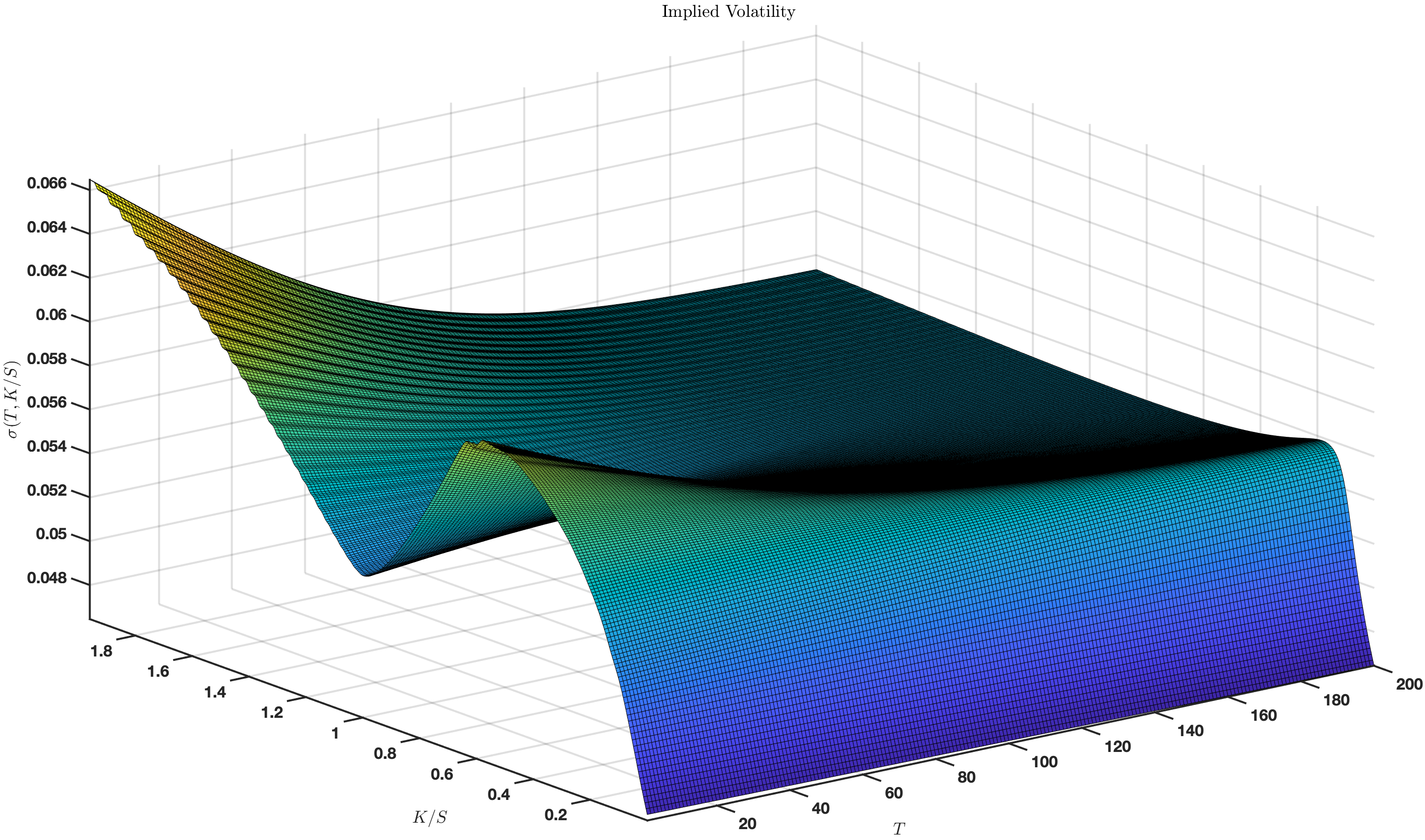}
\caption{}
\label{Call_implied_NDIG}
\end{subfigure}
\captionsetup{labelfont=bf,labelsep=quad}
\caption{NDIG-based price surfaces for bitcoin (a) call and (b) put options as functions of time to maturity and strike price.
(c) Implied volatility surfaces as functions of time to maturity and moneyness.}
\label{option_surfaces}
\end{center}
\end{figure}

Figures~\ref{option_surfaces} (a) and (b)  show the resulting prices for call and put options plotted against maturity, $\tau$,
and strike price $K$.
Figures (c) and (d) show the implied volatility surfaces, i.e., the market’s view on future volatility,
against time to maturity and relative moneyness, defined as $K/S$, where $S$ is the asset price obtained by the BSM model.
Figure~\ref{option_surfaces}-(c) investigate the behavior of implied volatility concerning moneyness and time to maturity for both call and put options. It is observed that as moneyness increases, moving towards out-of-the-money values, the implied volatility also increases. Conversely, as moneyness decreases, a distinct pattern emerges, referred to as the "smirk" in the implied volatility curve. Initially, as moneyness decreases and options move in-the-money, implied volatility rises, but subsequently, it starts to decrease for further in-the-money options, eventually approaching near-zero values. As we extend the time to maturity (denoted as $\tau$) of the options, we notice a noteworthy trend. The sensitivity of implied volatility to changes in moneyness diminishes for call options, and this effect appears to converge to a constant level irrespective of the moneyness. In other words, as the options approach their expiration dates, their implied volatility becomes less responsive to changes in moneyness.

\section{ Bitcoin Volatility Measures}\label{BitCoin_Vol}
\noindent
Understanding the volatility of speculative assets is critical for investment decisions.
Given that bitcoin is, at least by some, considered to be a potential alternative to fiat money,
volatility characteristics are of especial concern.
It is, therefore, of interest to understand and adequately model the process that governs the volatility of bitcoin.

Model parameters reflecting the volatility of asset prices are known to vary stochastically and exhibit clustering.
To allow for these phenomena in option pricing, \cite{hull1987pricing} and \cite{heston1993closed} suitably randomize
the volatility parameter in the Black–Scholes model, where the volatility process is governed by Brownian motion.
An alternative strategy was adopted by \cite{carr2003stochastic} based upon ideas by \cite{geman2001time} and \cite{Clark_1973}.
Clark conjectured that price processes are controlled by a random clock, which is a cumulative measure of economic activity,
and used transaction volume as a proxy for this measure.
Geman et al. suggested the price process must have a jump component;
thus the price process can be regarded as Brownian motion subordinated to this random clock.
Carr et al. developed this subordinated model using normal inverse Gaussian and variance gamma examples of pure jump L{\'e}vy processes.
\cite{hurst1997subordinated} considered various subordinated processes to model the leptokurtic characteristics of stock--index returns.
In the option pricing literature, Carr and Wu (2004) extended the approach in \cite{carr2003stochastic} by providing an efficient way to
allow for correlation between the stock price process and random time changes.
\cite{klingler2013option} introduced two new six--parameter processes based on time changing tempered stable distributions
and developed an option pricing model based on these processes.
\cite{shirvani2021multiple} introduced the volatility intrinsic time or volatility subordinator model to reflect the heavy--tail phenomena
present in asset returns.
They studied the question of whether the VIX index is a volatility index that adequately reflects intrinsic time. 
They showed that this index fails to properly capture the intrinsic time for the SPDR S\&P 500.
Apparently, the VIX index, as a measure of time change, does not reflect all the information needed to correctly capture the skewness
and the fat--tailedness of the S\&P 500 index.
A model with a suitable volatility subordinator should adequately account for such empirical phenomena.

In this section, we apply three views to analyze bitcoin volatility.
The first measures historical realized volatility through sample standard deviation.
As noted in the Introduction, this is the method employed in the Bitcoin Volatility Index.
The second measures implied (future) volatility in the risk-neutral, derivatives world, $\mathbb{Q}$,
reflecting the views of the option traders.
The third measures implied volatility in the real world, $\mathbb{P}$, reflecting the views of spot traders.
We employ an intrinsic time formulation in ths latter case.
In the following, we empirically compare these three modeling strategies using daily bitcoin data from
January 1, 2015 to April 8, 2021.

\begin{figure}[hbt!]
\begin{center}
\includegraphics[width = \textwidth]{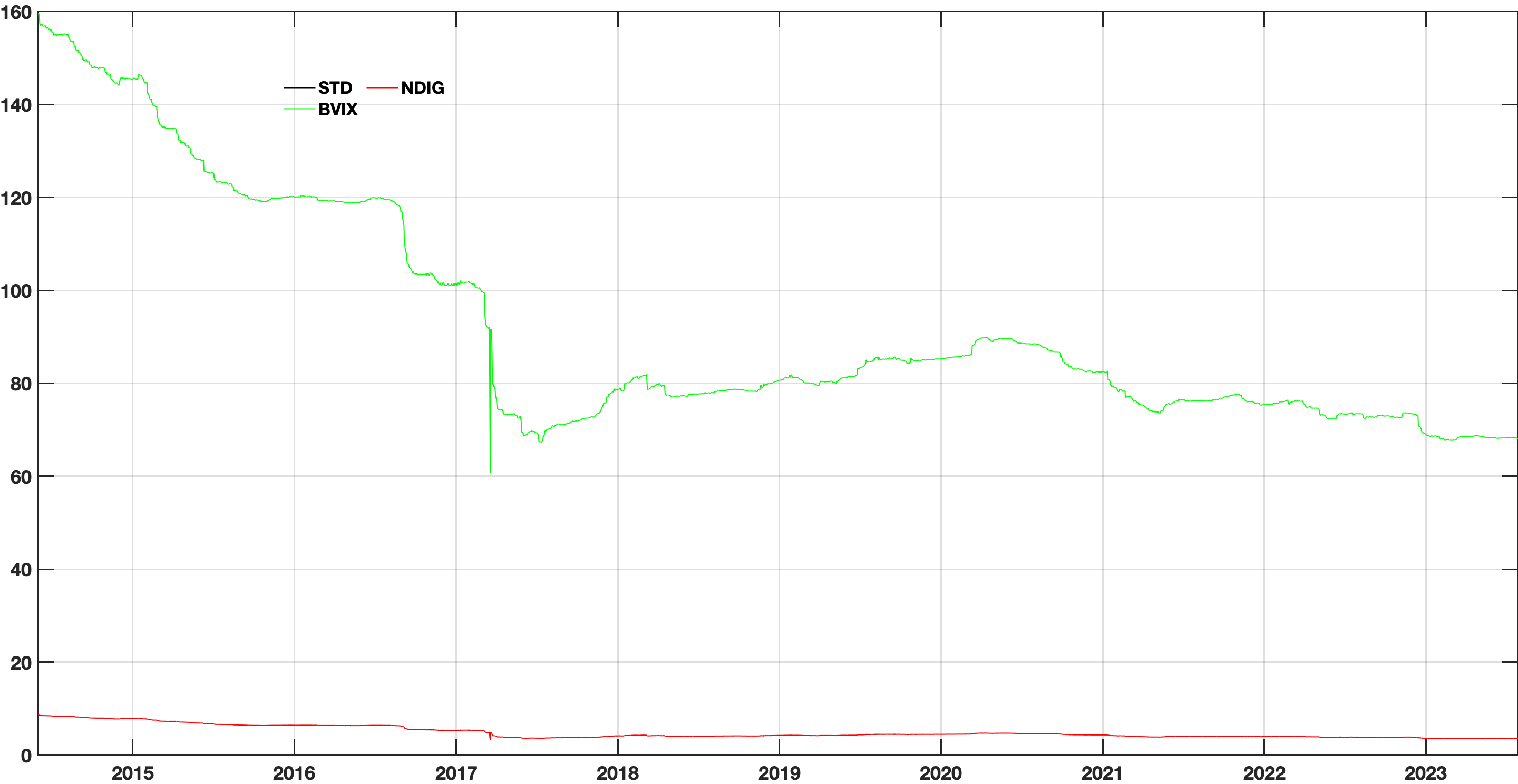}
\captionsetup{labelfont=bf,labelsep=quad}
\caption{Comparison of bitcoin volatilites computed historically (STD), from option prices (BVIX), and using intrinsic time (NDIG IT).
(Y-axis values are in percent.)}
\label{Vol_fits}
\end{center}
\end{figure}

\medskip
{\bf Historical Volatility}.
Figure~\ref{Vol_fits} plots the historical volatility series based on measuring average standard deviation on a 1008--day rolling window of bitcoin returns.
The historical volatility varies substantially over the sample period, assuming values between $60.69$ and $159.62$ percent.

\medskip
{\bf Implied Volatility in $\mathbb{Q}$}.
Option traders commonly use implied volatility as a proxy for future volatility.
The VIX, the index reflecting volatility expectations for the  S\&P 500 stock index, is based on S\&P 500 index options
estimated on a real-time basis by the Cboe.
The VIX can be regarded as an efficient volatility forecast for the  S\&P 500 index, provided that option markets are efficient.

\noindent The value of the VIX index is
\begin{equation}
VIX = 100 \times \sqrt{{W_1 \sigma_1^2 + W_2 \sigma_2^2}}
\label{eq:vix}
\end{equation}
Subscript “1” denotes near-term options while “2” denotes next term options. As SPX options expire on Fridays, depending on the trading day, near-term options have 23 to 30 days until expiration, while next-term options have 31 to 37 days until expiration. The weights $W_j$ $(j = 1,2)$ express these expiration times in a normalized form, accurate to the minute,
\begin{equation}
W_1 = \frac{M_{T_1}}{M_{30}} \left(\frac{ M_{T_2} - M_{30} }{M_{T_2} - M_{T_1}}\right), \quad
W_2 = \frac{M_{T_2}}{M_{30}} \left(\frac{ M_{30} - M_{T_1} }{M_{T_2} - M_{T_1}}\right),
\label{eq:weights}
\end{equation}
where $M_j$ represents the number of minutes until the settlement of the $j$-th term options, while $M_{30}$ represents the total number of minutes in a 30-day period. As a result, $W_1$ and $W_2$ are constrained to the range $0 \leq W_1,W_2 \leq 1$, with the additional requirement that $W_1 + W_2 = 1$. 
The equation below, provided by \cite{demeterfi1999guide}, can be used to calculate the near-term and next-term volatilities:

\begin{equation}
\label{next_term_vol}
\sigma^2_{j} = \frac{2e^{r_f T_j}}{T_j} \sum_{i}
\frac{\Delta K_i}{K_i^2} Q(K_i)
\frac{1}{T_j} \left( \frac{F_j}{K_{0j}} -1 \right)^2.
\end{equation}
Here, $T_j$ is the time to expiration measured in fractions of a year, $r_f$ is the annualized risk-free rate, $F_j$ is the forward index level derived from index option prices, $K_{0j}$ is the first strike price below $F_j$, $K_i$ is the strike price of the $i$th out-of-the-money option (a call if $K_i>K_0$, a put if $K_i<K_0$, and both call and put if $K_i=K_0$), $\Delta K_i$ is the interval between strike prices, and $Q_{K_i}$ is the midpoint of the bid-ask spread for each option with strike price $K_i$. For more information on the computation of each of these parameters, refer to the Cboe white paper by \cite{VIX_CBOE2019} on VIX.

To calculate a "VIX-like" volatility known as BVIX, the NDIG model is used to compute prices for European put and call options that have between 23 and 37 days to expiration. To compare with the historical volatility, BVIX values are calculated using historical data as follows: for each moving window of length 1008 Bitcoin log-return data,

\begin{enumerate}
    \item  The NDIG model is fitted to the 1008 Bitcoin log-return data (4 years), and the model parameters are estimated using the minimization \ref{eq:Example1}. The resulting parameter values are displayed in Fig. \ref{Parameters}. 
The results depicted unveil discernible patterns among the parameters under investigation. Particularly noteworthy is the conspicuous observation that $\lambda_T$ displays the least temporal variation, while $\rho$ showcases the highest level of temporal variability.
Furthermore, an examination of the estimated values for $\mu_3$ and $\sigma_3$ indicates a declining trend leading up to April 2017, followed by a subsequent abrupt change, with relatively minor fluctuations thereafter.
Additionally, it is worth noting that the parameter $\rho$ exhibits the smallest magnitude in its estimated value, accompanied by a considerable degree of dispersion or deviation from the mean. These observations are graphically represented in Figure \ref{Parameters}, which presents a comparative visualization of the first four empirical moments derived from the time series depicted in Figure \ref{Parameters} . These moments are contrasted with those computed from the parameters derived via Equation \ref{eq:Example1}. This analysis offers valuable insights into the dynamics and characteristics of the examined parameters.

Fig. \ref{Moments} compares the first four
empirical moments computed from the time series in Fig. \ref{Fig_1} and the moments computed from
the fitted parameters using \ref{eq_moments}.

\begin{figure}[h]
\begin{center}
\begin{subfigure}[b]{0.49\textwidth}
\includegraphics[width=\textwidth]{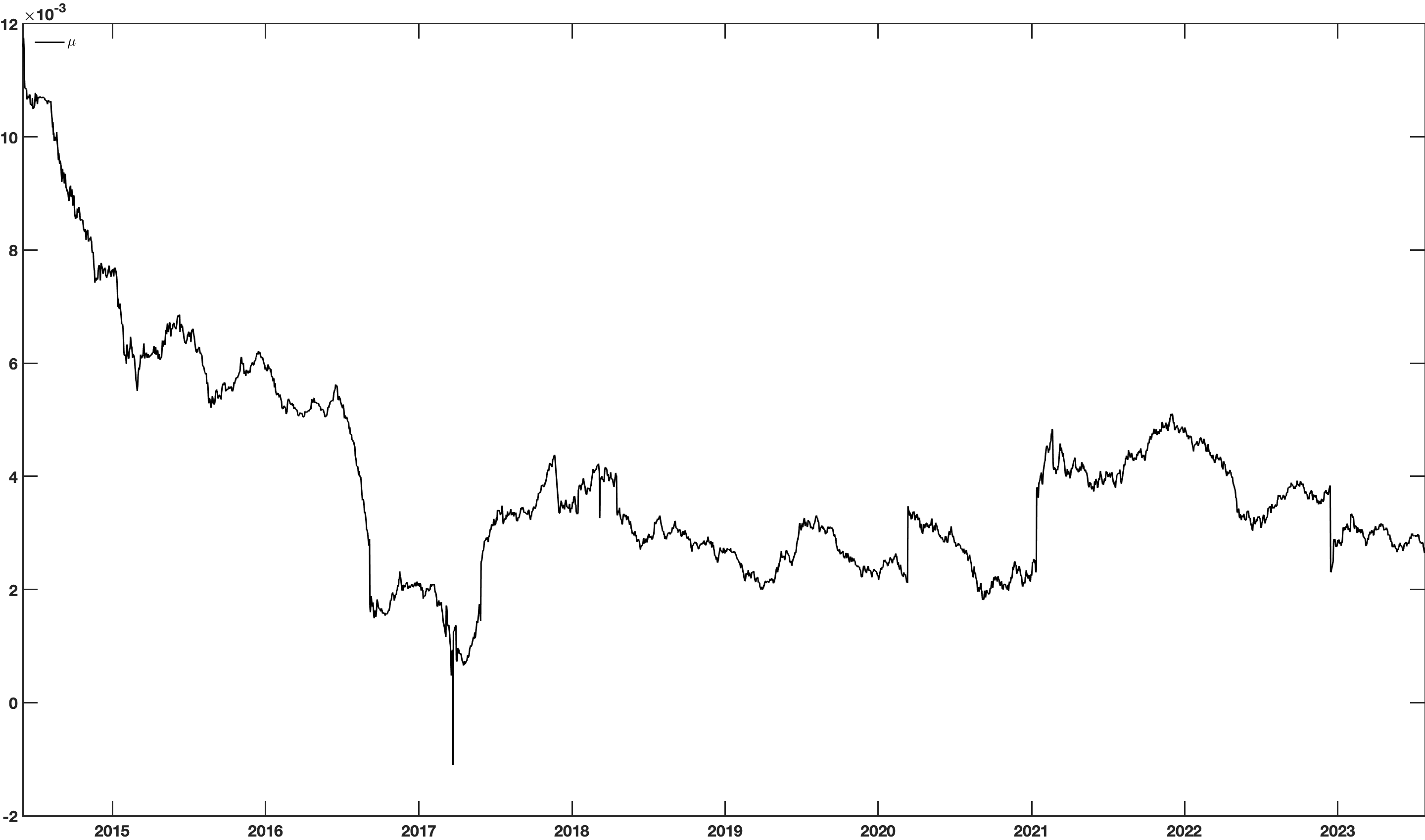}
\caption{}
\label{Mu}
\end{subfigure}
\begin{subfigure}[b]{0.49\textwidth}
\includegraphics[width=\textwidth]{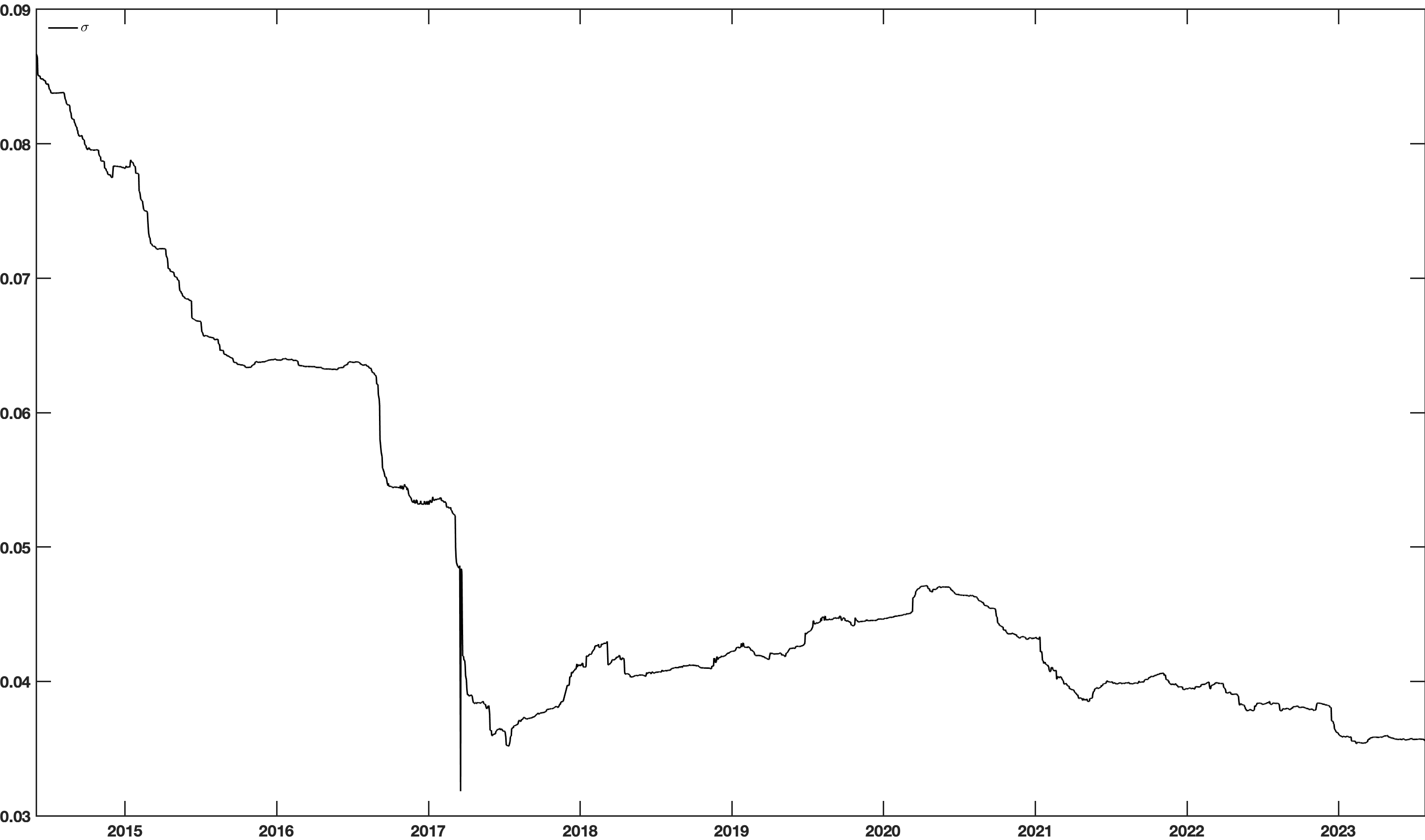}
\caption{}
\label{Sigma}
\end{subfigure}
\begin{subfigure}[b]{0.49\textwidth}
\includegraphics[width=\textwidth]{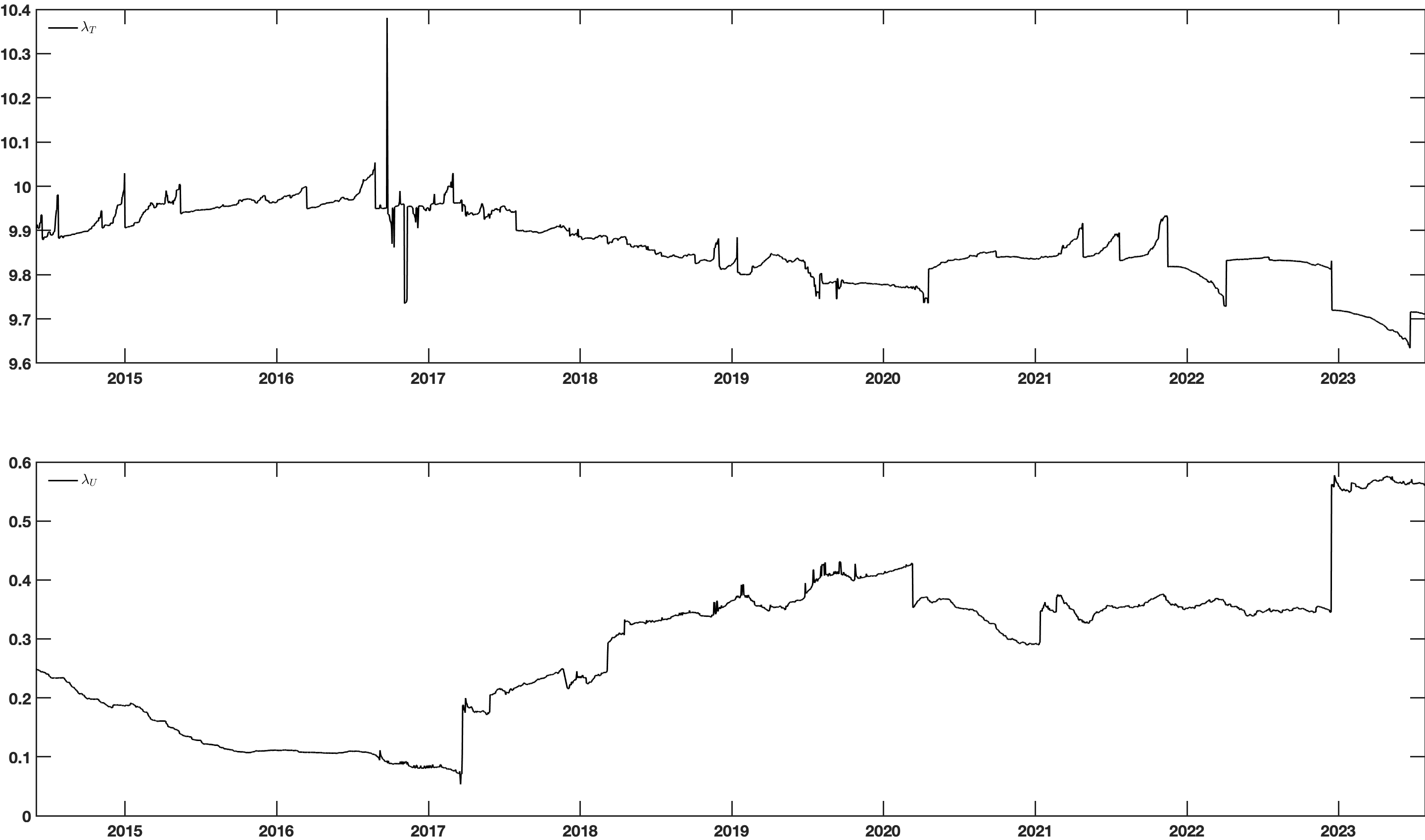}
\caption{}
\label{Lambda}
\end{subfigure}
\begin{subfigure}[b]{0.49\textwidth}
\includegraphics[width=\textwidth]{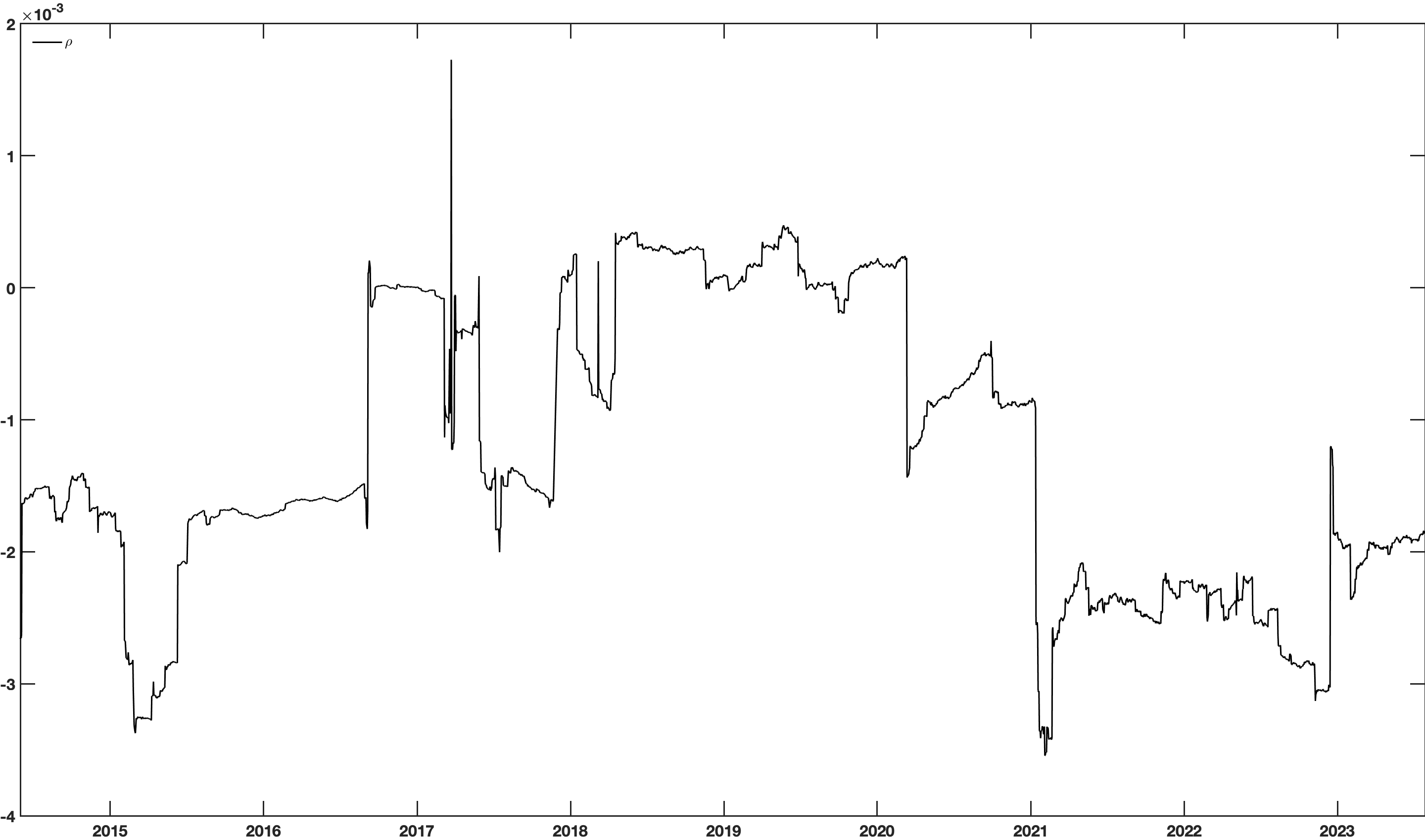}
\caption{}
\label{Rho}
\end{subfigure}
\captionsetup{labelfont=bf,labelsep=quad}
\caption{Four-year moving window fits to the parameter values (a) $\mu_3$, (b) $\sigma_3$, (c) $\lambda_T$, and $\lambda_U$, and (d) $\rho$.}
\label{Parameters}
\end{center}
\end{figure}

\begin{figure}[h]
\begin{center}
\begin{subfigure}[b]{0.49\textwidth}
\includegraphics[width=\textwidth]{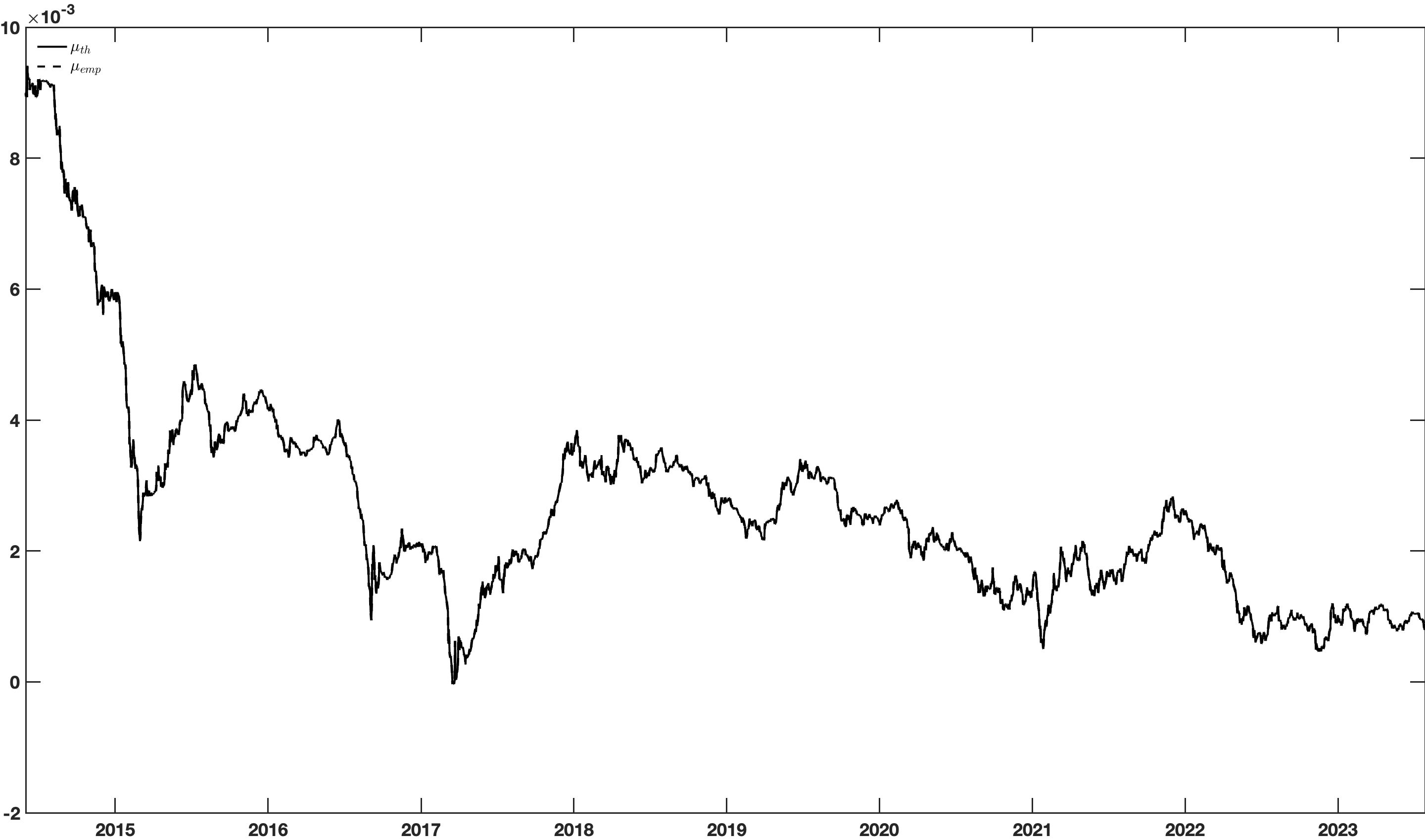}
\caption{}
\label{Mean}
\end{subfigure}
\begin{subfigure}[b]{0.49\textwidth}
\includegraphics[width=\textwidth]{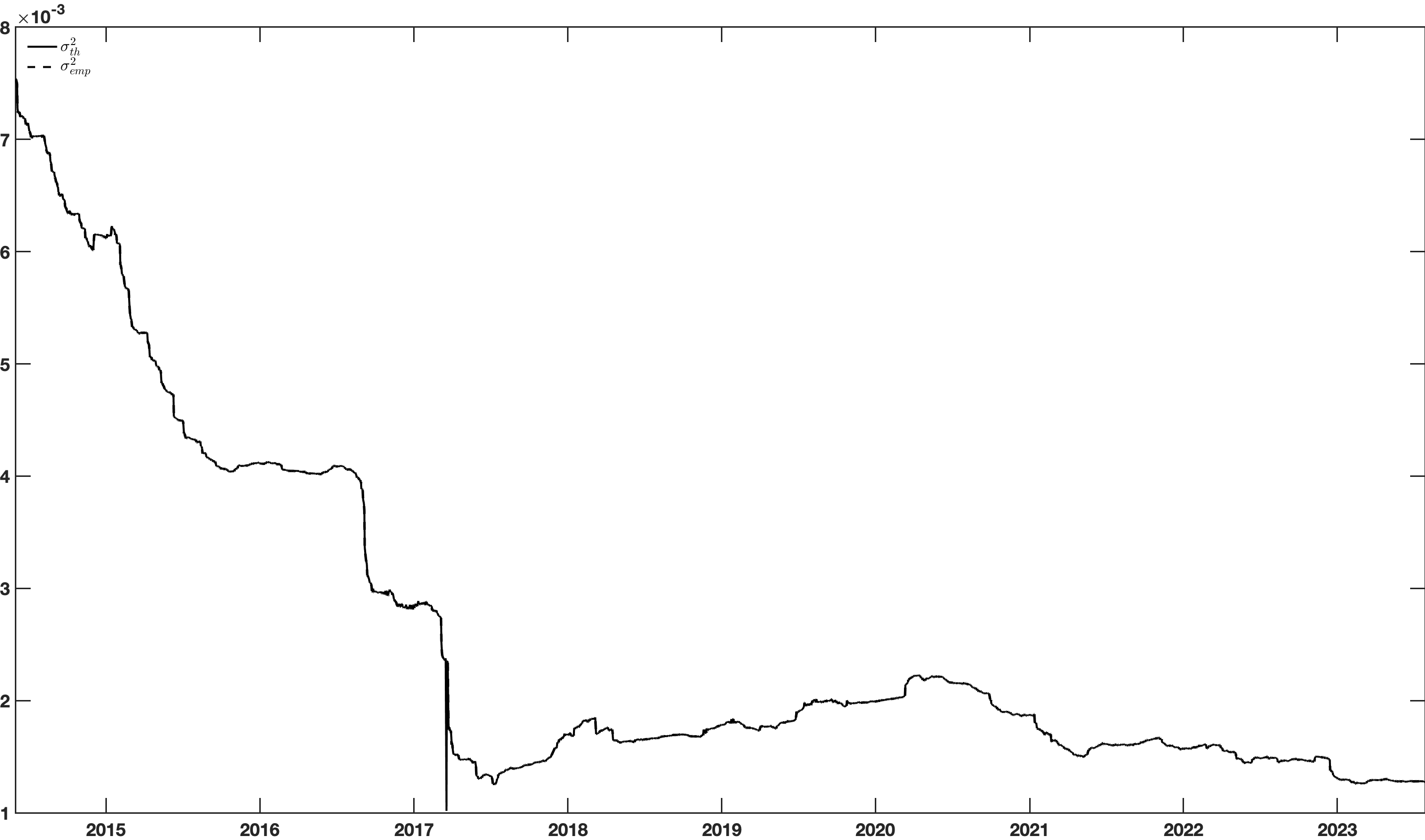}
\caption{}
\label{Variance}
\end{subfigure}
\begin{subfigure}[b]{0.49\textwidth}
\includegraphics[width=\textwidth]{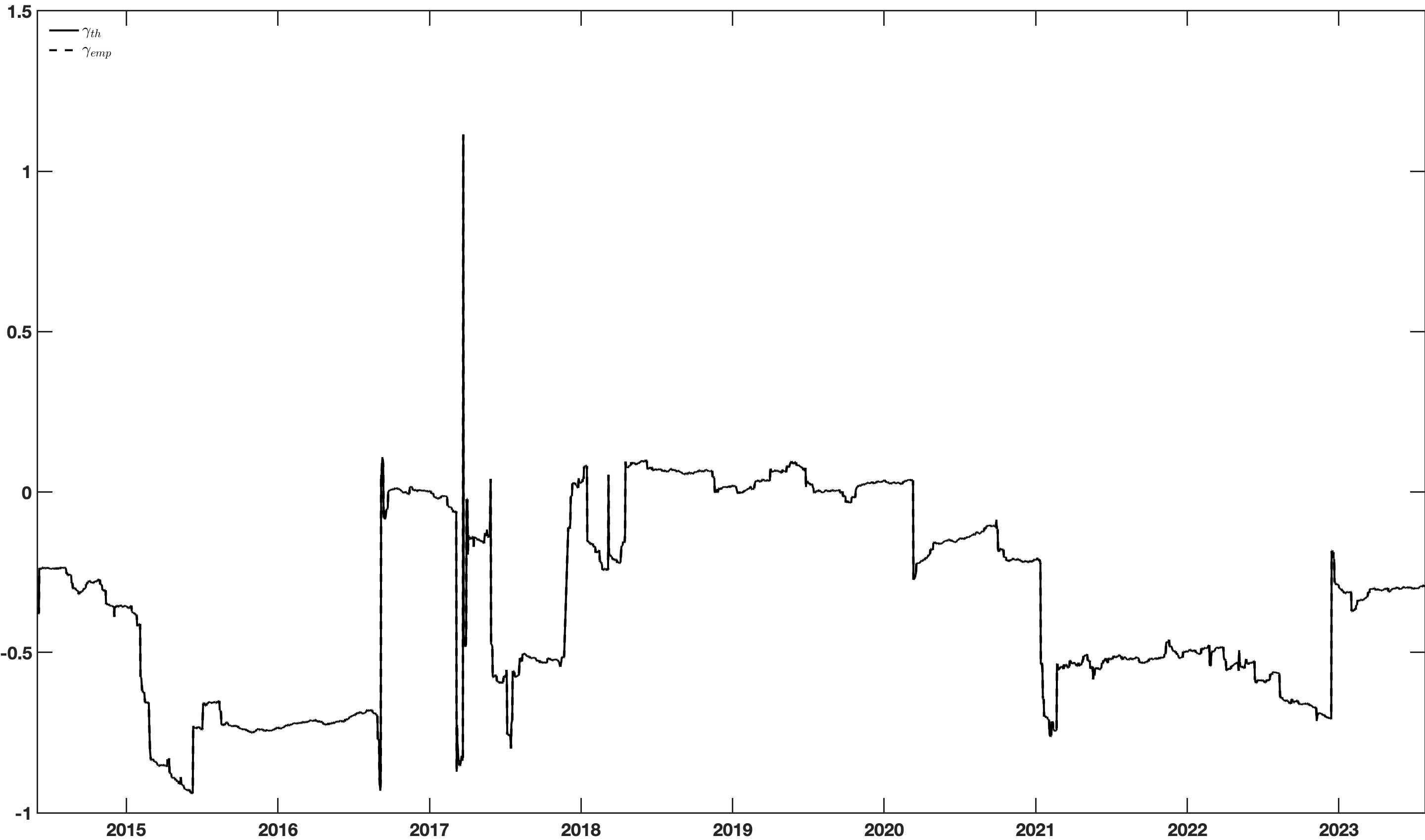}
\caption{}
\label{Skewness}
\end{subfigure}
\begin{subfigure}[b]{0.49\textwidth}
\includegraphics[width=\textwidth]{Variance.png}
\caption{}
\label{Variance}
\end{subfigure}
\captionsetup{labelfont=bf,labelsep=quad}
\caption{ Comparison of the first four empirical (“emp”) moments computed from the return time series and the moments (“th”) computed from the fitted parameters.}
\label{Moments}
\end{center}
\end{figure}

  \item The dynamics of the Bitcoin price $S_T^\mathbb{Q}$ in the risk-neutral world and subsequently, the characteristic function (CF) of the log price $\ln S_T^\mathbb{Q}$ are determined using \ref{chf_Log_price}.

  \item Call option prices with appropriate expiration dates are computed using the FFT formulation \ref{Carr_call}, and put option prices are computed using put-call parity based on the portfolio spot price $S_T^\mathbb{Q}$ for the last date of the moving window. In the context of option price calculations based on \ref{Carr_call}, the numerical values for the parameters $a$, $a_{\text{max}}$, and $v_{\text{max}}$ are determined through the utilization of conditions outlined in  \ref{a_condition_final} and Equation \ref{vmax}, respectively. Choosing too small values for $a$ 
 results in option prices that exceed the
maximum threshold, while choosing $a$ too large produces prices that go below the minimum
threshold (and maybe negative). As we discussed, a stricter determination for the value of $a \in [0, a_{max}]$ is based on the requirement that option prices remain within allowed bounds.  Fig. \ref{amax} shows  the computed values for the upper bound $a_{max}$  over period 05/08/2014 through
07/28/2023. For the computations in the remainder of this
section, we utilize the smallest possible value, $a =  0.40$.

\begin{figure}[hbt!]
\begin{center}
\includegraphics[width = 0.8\textwidth]{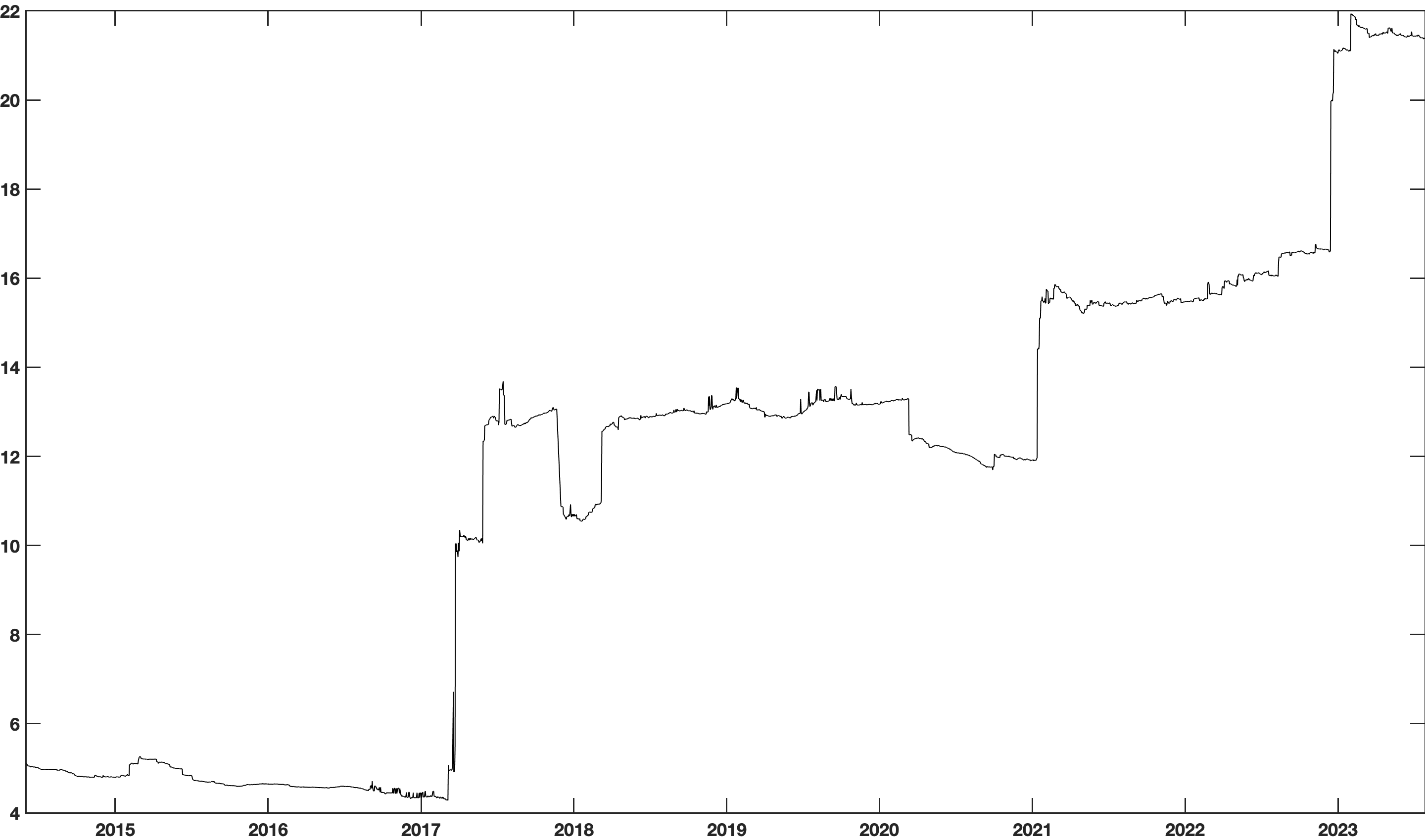}
\captionsetup{labelfont=bf,labelsep=quad}
\caption{Values for the upper bound $a_{max}$ computed over the period 05/08/2014 through 07/28/2023.}
\label{amax}
\end{center}
\end{figure}

  \item BVIX values are then computed using \ref{eq:vix} through \ref{next_term_vol}. As there are no traded options on the Bitcoin data, the following minor modifications were made to the VIX formulation:

\begin{enumerate}
    \item The risk-free interest rate used is the annualized bond-equivalent (coupon-equivalent) yield for 3-months U.S. treasury bills published for day $t$.

\item Closing times for option evaluation are at $4:00$ PM on day $t$.

\item Options expire on near- and next-term Fridays at $4:00$ PM.

\item As there is no bid-ask spread in the NDIG option price computations, $Q_{K_i}$ is computed directly as the NDIG option price.

\item The range of strike prices considered in the BVIX computation is from $0.75 S_t$ to $1.5 S$, as NDIG computed option prices do not go to zero.
\end{enumerate}
\end{enumerate}
As mentioned in step 4d, since the underlying asset portfolio is not traded, there is no market sentiment setting bid-ask prices for the call and put options used in the BVIX calculation. Therefore, the BVIX volatility directly reflects the NDIG option price computations, and there is no risk premium or market price of risk that is typically captured by the VIX.

Figure~\ref{Vol_fits} shows the values for BVIX.
As observed from this figure, BVIX  values are higher than STD values.
The fact that implied volatility computed from options based upon a risky asset is higher than realized volatility is commonly observed and
is referred to as the ``volatility risk premium.''
An explanation for the volatility risk premium is that option traders allow for the probability of a significant market move.
To be compensated for the inherent volatility risk that cannot be hedged away, option sellers charge a premium.
Option traders typically view implied volatility as the most reliable approach to volatility forecasting
\citep{jorion1995predicting}.
Figure~\ref{Vol_fits} and a Pearson correlation coefficient of $\rho = 0.48$ between BVIX and historical volatility
indicates that implied volatility provides only partial information about future empirical volatility.

\begin{figure}[hbt!]
\begin{center}
\includegraphics[width = \textwidth]{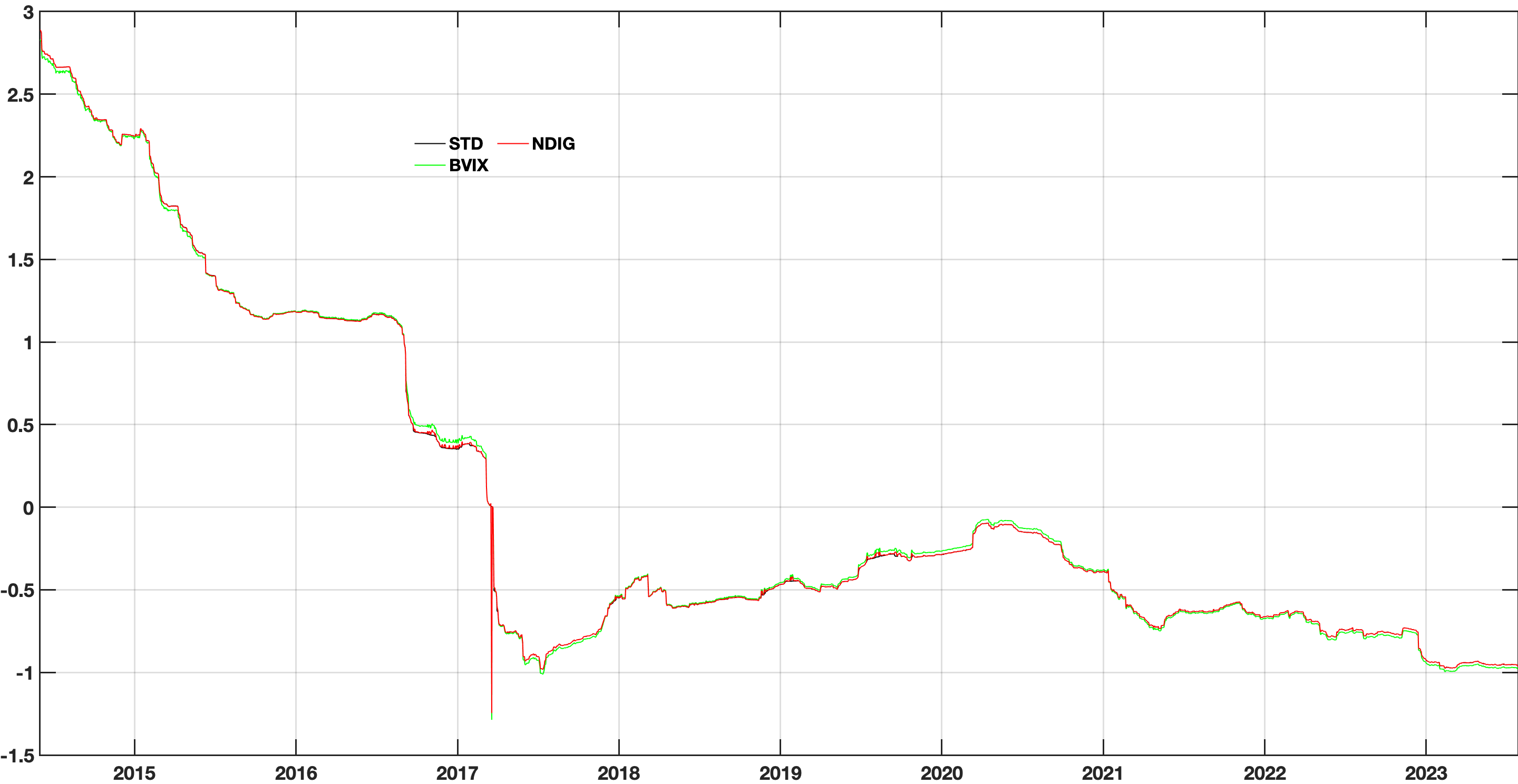}
\captionsetup{labelfont=bf,labelsep=quad}
\caption{Normalized (mean subtraced, standard deviation scaled) versions of the bitcoin volatility series in Figure~\ref{Vol_fits}.}
\label{Vol_fits_norm}
\end{center}
\end{figure}

To account for the different scales, the historical and BVIX volatility time series can be normalized separately by subtracting the series mean and dividing by the series standard deviation. These normalized plots, as shown in Fig. \ref{Vol_fits_norm}, demonstrate close agreement between the two, with the BVIX volatility displaying a slight daily fluctuation that is not present in the historical volatility. This emphasizes the fact that the sole source of variance in this numerical example is the daily returns of the underlying risky asset, the Bitcoin data, which is directly quantified by the historical volatility. Since all option prices in this example are calculated using the Carr-Madan formulation and the NDIG model, no additional volatility is introduced to these option prices through, for instance, trader sentiment (bid-ask pricing). Therefore, the BVIX volatility formulation only captures the original asset return volatility.

\medskip
{\bf Implied Volatility in $\mathbb{P}$: Intrinsic Time}.
\cite{Mandelbrot_1967} coined the term "intrinsic time" in finance to describe how market information is not continuously available, but instead arrives in discrete events occurring at varying intervals. Market orders for assets, for instance, are a prime example of such events, which can differ significantly in both timing and frequency throughout the trading day. These events provide information on asset value (price), and their occurrence, magnitude, and sign characterize intrinsic time. Price change information may also differ in its level of informativeness, with larger changes providing more information than smaller ones, and consecutive price changes of the same sign providing more information than those with opposite signs. Researchers have explored the concept of intrinsic time and applied it to financial time series in various works, including 
\cite{guillaume1997bird}, \cite{tsang2010directional}, and \cite{aloud2012directional}.

According to the intrinsic time perspective, no information is available between events, and therefore no time has elapsed. An analogy to conceptualize intrinsic time is to imagine an analog clock's seconds-hand that ticks intermittently and randomly, varying in volume with each tick. Time only moves forward when the hand advances, and the tick's volume represents the amount of information conveyed.

By utilizing the double subordinated method, a novel measure of volatility can be derived under the $\mathcal{P}$ measure.
From \eqref{DSS}, we view $X_t$ as the process governing the intrinsic time of the double subordinated price process.
From  \eqref{eq_moments}  with $\gamma = 0$, the standard deviation (i.e. the volatility) of the unit increment of $X_t$,
when $X_t$ follows a NDIG log--price process, is given by
\begin{equation}\label{Eq13}
\textrm{Vol} (X_1) \equiv \sqrt{\textrm{Var}(X_1)}
= {\sigma_3}^2 + \rho^2 \left( 
\frac{1}{\lambda_T}
+\frac{1}{\lambda_U}
\right)
\end{equation}
where $\lambda_T$, $\lambda_U$, and $\sigma_3$ are model parameters as defined in Table \ref{NDIG_Par}.
We can refer to the volatility $\sqrt{\textrm{Var}(X_1)}$ as the NDIG volatility, which takes into account both the Brownian motion (Gaussian) and L\'{e}vy subordinator components of the model. However, it is important to note that this measure differs from other measures of volatility in the literature.

To derive the intrinsic time for the bitcoin log--price process, we compute rolling NDIG parameter estimates from the recent six years
of daily data, using overlapping windows with a length of 1008 days.
We use \eqref{Eq13} to derive the annualized volatility implied by the NDIG model.
The volatility computed by the NDIG intrinsic time model (NDIG IT) is also shown in Figure~\ref{Vol_fits}.

Examination of all three volatility plots reveals substantial similarity between the patterns of NDIG intrinsic time and historical volatility,
confirming that \eqref{Eq13} captures realistic bitcoin dispersion patterns.
We have addressed above the observed differences between the BVIX implied volatility and  historical volatility.
We note that the NDIG IT volatility predictions are consistently larger than historical, but otherwise demonstrate very strong synchronicity.
Normalized${}^9$ versions of the three series are plotted in Figure~\ref{Vol_fits_norm}.
The normalized NDIG IT and STD time series are now virtually identical, indicating an almost perfect linear correlation between the two.
The BVIX variant is less synchronous but still co-moves to a certain degree,
with a strong pattern of anticipating cycles of high volatility.

\begin{table}[h]
\centering
\captionsetup{labelfont=bf,labelsep=quad}
\caption{$p$-values from augmented Dickey-Fuller (ADF), KPSS and Johansen tests}
\label{tab:pvals}
\begin{tabular}{lcccc l c c}
\toprule
Test & BVIX  & NDIG IT     & STD   & \ &  Test        & $r = 0$ & $r = 1$ \\
\midrule
ADF   & 0.009 & 0.007 & 0.001 & \ & Johansen  & 0.001 & 0.001  \\
KPSS & 0.010 & 0.010 & 0.010 & \ &\omit         & \omit & \omit \\
\omit & \omit & \omit  & \omit & \ & \omit        & \omit & \omit \\
\bottomrule
\end{tabular}
\end{table}

Cointegration analysis is one approach to assess whether or not there are significant deviations in co-movement among time series.
To apply cointegration, we first check whether each volatility series is integrated.
The results of both the augmented Dickey-Fuller and KPSS tests \citep{kwiatkowski1992testing} shown in 
Table \ref{tab:pvals} support the hypothesis that each series is integrated of order one (has a unit root).
Performing a Johansen test \citep{johansen1988statistical} involving all three time series (Table \ref{tab:pvals})
indicates a cointegrating relationship of orders 1 and 2 among the three time series.

Taken together, the results from Figure~\ref{Vol_fits_norm} and Table~\ref{tab:pvals} show that the dynamic NDIG IT model,
which is consistent with dynamic asset pricing theory and has predictive capabilities,
provides very good agreement when applied ``in-sample''.
The implied volatility model, BVIX, which is also dynamic and predictive, has poorer agreement with the in-sample data.
The pattern of BVIX often, though not always, anticipating cycles of high volatility may speak to option traders' heightened sensitivity
to potential increases in market volatility.

\begin{figure}[hbt]
\centering
\includegraphics[width = 0.75\textwidth]{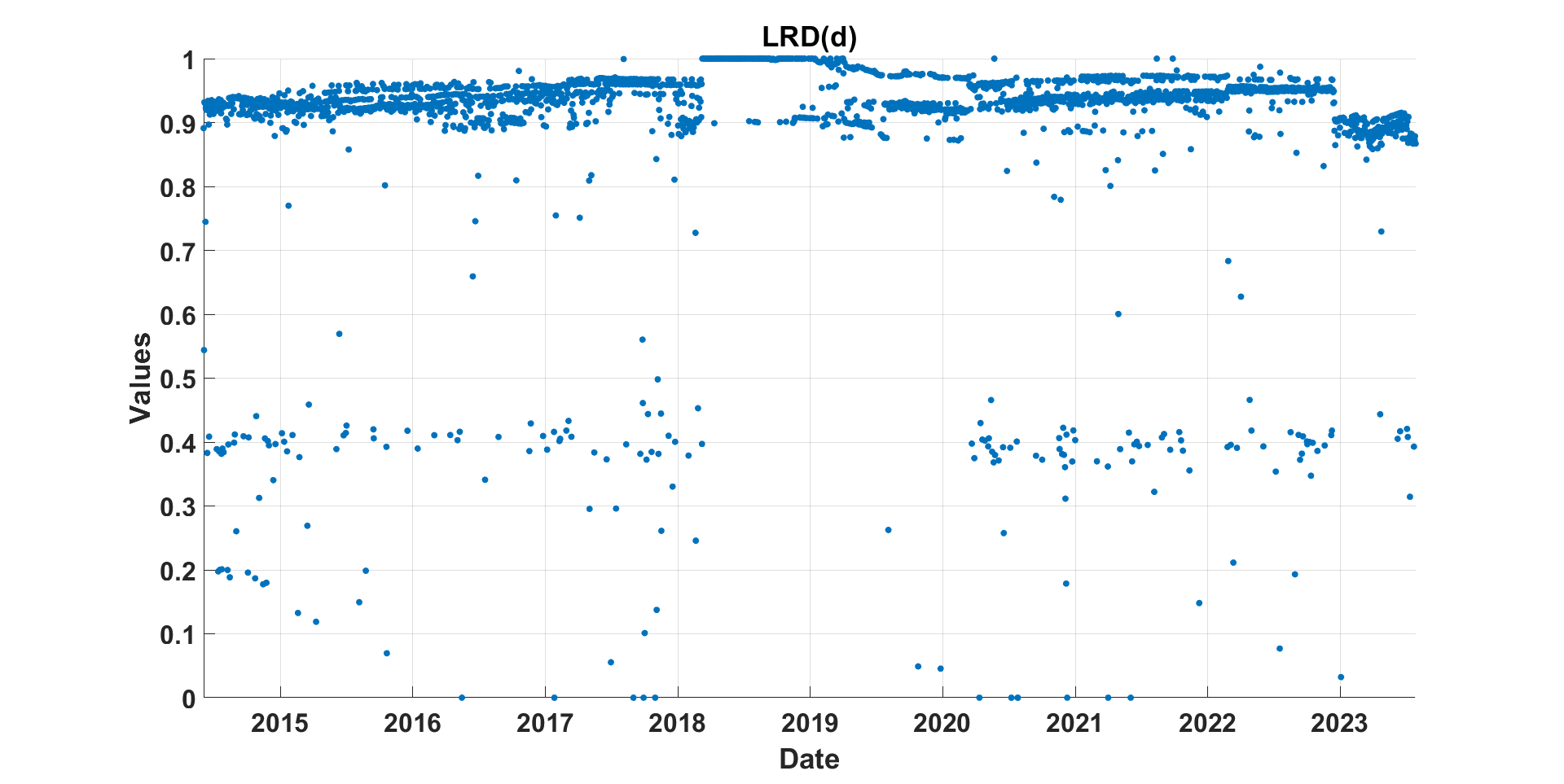}
\captionsetup{labelfont=bf,labelsep=quad}
\caption{Time-varying LRD $(d)$ in the bitcoin volatility log--return process.}
\label{LRD_VOl}
\end{figure}

Finally, we investigate the question of whether there is long--range dependency (LRD) in bitcoin volatility, as has been found
in stock volatility \cite[see, for example,][]{asai2012asymmetry}.
We fit a fractionally integrated autoregressive conditional heteroskedastic (FIGARCH) model to investigate both short-- and long--range dependency in the bitcoin variance series.
Specifically, we fit an AR(1)-FIGARCH(1,$d$,1)${}^{10}$ with Student's $t$ innovations.
Fractional integration in the GARCH part, i.e., $d>0$, indicates LRD in volatility and non-stationarity.
For $d<0$ we have anti-persistency and short--range memory.
The AR(1)-FIGARCH(1,$d$,1) model parameters are estimated in a rolling fashion using a window of 252-day window.
Figure~\ref{LRD_VOl} plots the sequence of $d$ estimates in the conditional heteroskedastic equation for the variance.
The estimates fall almost exclusively in the interval $(0.8,1.0)$, indicating non--stationarity and the presence of
LRD in bitcoin volatility.
This implies that volatility measures derived from historical data can be used to determine bitcoin volatility---a
finding that also follows from Figure~\ref{Vol_fits_norm},
showing that the (normalized) historical standard deviation and NDIG intrinsic time move closely together.

\section{\normalsize Discussion}\label{Sec_Discussion}
\noindent

Dealing with heavy--tailed asset returns remains a challenge in theoretical and empirical finance.
Implicit in our approach is that there is a relationship between arrival times of financial news, the ``intrinsic time'' of finance,
and price volatility for an asset, and particularly for bitcoin, whose volatility is especially pronounced.
As noted by \cite{Clark_1973},
``The different evolution of price series on different days is due to the fact that information is available to
traders at a varying rate.
On days when no new information is available, trading is slow, and the price process evolves slowly.
On days when new information violates old expectations, trading is brisk, and the price process evolves much faster.''
Our analysis is based on the assumption that information (whether unexpected or reinforcing) arrives at a rate
that defines an ``intrinsic'' or ``operational'' time scale.
Whether this intrinsic time is measured in rate (ticks) or volumes of transactions, or another proxy is irrelevant;
we take as a stylized fact that the intrinsic time distribution of information arrival is skewed and extremely heavy tailed.
(See, for example, \cite{Chakraborti_2009} on the microstructure of the time between successive transactions.)

The philosophy behind our approach is therefore as follows.
The log-price process is additive Brownian motion (Equation \eqref{Black_Eq2}),
but the time variable that drives the Brownian motion is not the physical time $t$ as suggested by \eqref{Black_Eq2},
but is an intrinsic or operational time.
We utilize subordinated L{\'e}vy processes to mimic a relationship between intrinsic time and price processes observed in
physical time.
Unfortunately, as has been demonstrated by \cite{Shirvani_2021}, a single level of subordination (equation \eqref{Eq2}) is insufficient
to capture both the stylized facts of the operational time scale and the stylized facts (skewness, heavy-tails, clustering) observed in
asset returns.
In this paper we show that, for a very volatile asset, a doubly subordinated  process (equation \eqref{DSX})
using a normal inverse Gaussian process at each level of subordination
is capable of capturing (Figure~\ref{Ksdensity_NDIG})
the skewness and heavy-tailed behavior of the asset's return process in physical time,
and is, presumably also accounting for the heavy-tailed distribution of the underlying operational time.

By extending our double subordination approach to option pricing, we are able to show convincingly that
not only does the double subordination model capture the distributional properties of the price dynamics,
it very accurately captures (Figure~\ref{Vol_fits}) the trend of the historical (in-sample) bitcoin volatility
(as measured by historical standard deviation).
Further, Figure~\ref{Vol_fits_norm}, demonstrates that we can derive the correct linear scaling to be
applied to the double  subordinated model to capture actual (in-sample) volatility.
\nc

\bigskip
\noindent\textsc{Notes}

\noindent
${}^1$ See \url{https://www.buybitcoinworldwide.com/volatility-index/}.

\noindent
${}^2$ The CRIX index captures the performance of a basket of ten leading cryptocurrencies that are listed on well--known,
transparent exchanges and meet certain liquidity and market capitalization requirements.
S\&PGlobal, through its S\&P Dow Jones Indices division, has also introduced an index,
the S\&P Cryptocurrency MegaCap Index \citep[see][]{Cryptocurrency_index},
which is comprised of the S\&P Bitcoin Index and the S\&P Ethereum Index.

\noindent
${}^3$ See \cite{Bochner_1955}, \cite{Sato_1999} and \cite{Schoutens_2003}.

\noindent
${}^4$ For a further discussion of the use of subordinators in financial modeling, see \cite{Sato_1999} and \cite{Schoutens_2003}.

\noindent
${}^5$ We use ${\sim}$ to denote ``equal in distribution'' or ``equal in probability law.''

\noindent
${}^6$ As reported by \textit{Bloomberg Professional Services}, access date 04-08-2021.

\noindent
${}^7$ See \citet[][Chapter 6]{duffie2010dynamic}.

\noindent
${}^8$ This can be adapted in a straightforward manner for higher order integration rules.

\noindent
${}^9$ Specifically, the data set $X = \{ x_1, x_2, \dots, x_n\}$ is normalized as $\bar{X} = \{ \bar{x}_1, \bar{x}_2, \dots, \bar{x}_n\}$,
with $\bar{x}_i = (x_i - x_{\textrm{min}})/s_x$,
where $x_{\textrm{min}} = \min_{i=1}^n (x_i)$ and $s_x$ is the standard error of the data set.

\noindent
${}^{10}$ Cf.\ \cite{granger1980introduction}, \cite{engle1982autoregressive} and \cite{bollerslev1996modeling}.

\clearpage
%\newpage
\normalem

\iffalse
\bibliographystyle{chicago}
\bibliography{ref}
\fi

\end{document}